\documentclass[12pt]{article}%
\usepackage[nosort]{cite}
\usepackage{graphicx}
\usepackage{multicol}
\usepackage{amsfonts}
\usepackage{amssymb}
\usepackage{amsmath}
\usepackage{dsfont}
\usepackage{heck}
\usepackage{afterpage}
\usepackage{setspace}
\usepackage{verbatim}
\usepackage{color}
\usepackage{longtable}
\usepackage{float}
\usepackage{subcaption}
\usepackage{epsfig}
\usepackage{epstopdf}
\usepackage{adjustbox}
\usepackage{todonotes}

\usepackage{tikz}
\usepackage[margin=1in]{geometry}
\usepackage{titletoc}
\usepackage{hyperref}%
\usepackage{mathdots}
\providecommand{\U}[1]{\protect\rule{.1in}{.1in}}
\pdfoutput=1
\newsavebox{\mysavebox}

\usetikzlibrary[shapes.geometric]
\usetikzlibrary{positioning}
\usetikzlibrary{calc,intersections,through,backgrounds}
\tikzset{>=stealth}
\usetikzlibrary{decorations.pathmorphing}
\usetikzlibrary{decorations.markings}
\tikzset{>=stealth}
\hypersetup{colorlinks,citecolor=black,filecolor=black,linkcolor=black,urlcolor=black,pdftex}
\usetikzlibrary{decorations.markings}

\numberwithin{equation}{section}

\newcommand{\ba}{\begin{eqnarray}}
\newcommand{\ea}{\end{eqnarray}}

\newcommand{\be}{\begin{equation}}
\newcommand{\ee}{\end{equation}}

\tikzstyle{startstop} = [rectangle, rounded corners, minimum width=3cm, minimum height=1cm,text centered, draw=black, fill=blue!10]
\tikzstyle{startstop} = [rectangle, rounded corners, minimum width=3cm, minimum height=1cm,text centered, draw=black, fill=blue!10]
\tikzstyle{io} = [trapezium, trapezium left angle=70, trapezium right angle=110, minimum width=3cm, minimum height=1cm, text centered, draw=black, fill=blue!30]
\tikzstyle{process} = [rectangle, minimum width=3cm, minimum height=1cm, text centered, draw=black, fill=orange!30]
\tikzstyle{decision} = [diamond, minimum width=3cm, minimum height=1cm, text centered, draw=black, fill=green!30]
\tikzstyle{arrow} = [thick,->,>=stealth]
\begin{document}

\date{August 2017}

\title{6D Fractional Quantum Hall Effect}

\institution{PENN}{\centerline{${}^{1}$Department of Physics and Astronomy, University of Pennsylvania, Philadelphia, PA 19104, USA}}

\institution{Uppsala}{\centerline{${}^{2}$Department of Physics and Astronomy, Uppsala University, Box 516, SE-75120 Uppsala, Sweden}}

\authors{Jonathan J. Heckman\worksat{\PENN}\footnote{e-mail: {\tt jheckman@sas.upenn.edu}}
and Luigi Tizzano\worksat{\Uppsala}\footnote{e-mail: {\tt luigi.tizzano@physics.uu.se}}}

\abstract{We present a 6D generalization of the fractional quantum Hall effect
involving membranes coupled to a three-form potential in the presence of a large background
four-form flux. The low energy physics is governed by a bulk 7D topological field theory of abelian
three-form potentials with a single derivative Chern-Simons-like action coupled to
a 6D anti-chiral theory of Euclidean effective strings. We derive the fractional
conductivity, and explain how continued fractions which figure prominently in the classification of
6D superconformal field theories correspond to a hierarchy of excited states.
Using methods from conformal field theory we also compute the analog of the Laughlin wavefunction.
Compactification of the 7D theory provides a uniform perspective
on various lower-dimensional gapped systems coupled to boundary degrees of freedom. We also show that
a supersymmetric version of the 7D theory embeds in M-theory, and can be decoupled from gravity.
Encouraged by this, we present a conjecture in which
IIB string theory is an edge mode of a $10+2$-dimensional bulk topological theory, thus
placing all twelve dimensions of F-theory on a physical footing.}

\maketitle

\setcounter{tocdepth}{2}

\tableofcontents


\newpage

\section{Introduction}

Extra dimensions provide a unifying perspective on a variety of
lower-dimensional phenomena. This is by now quite commonplace in developing
connections between the higher-dimensional world of string theory and various
low energy effective field theories. It also figures prominently in the
analysis of many condensed matter systems, especially in the context of a bulk
theory in a gapped phase coupled to boundary modes.

A classic example is the 2D fractional quantum Hall effect (FQHE) \cite{Tsui:1982, Laughlin:1983fy,
Haldane:1983xm, Halperin:1983zz, Halperin:1984fn, Jain:1989tx} and its
connection with a bulk 2+1-dimensional Chern-Simons
theory \cite{Girvin:1987fp, Wen:1989na, Blok:1990mc, Wen:1990zza, Ezawa:1991,
Wen:1991, Frohlich:1991wb, Wen:1992vi, Zee:1996fe}. This
involves an interplay between the theory of 2D chiral conformal
field theories (CFTs), and a bulk topological field theory \cite{Moore:1991ks}.
For a recent introduction to the subject see \cite{Tong:2016kpv}. Various
generalizations of this phenomena are by now available, including a
higher-dimensional (integer)\ quantum Hall effect \cite{Zhang:2001xs}. For
additional discussion of the higher-dimensional quantum Hall effect in the
condensed matter and string theory literature, see e.g. \cite{Bernevig:2000yz,
Fabinger:2002bk, Bernevig:2002eq, Karabali:2002im, Bernevig:2003yz, Karabali:2003bt, Hasebe:2016tjg}. There is by now a vast literature
on various ways that bulk gapped systems produce novel edge mode dynamics.\footnote{The list of
such references is well-known to experts. For a review of some aspects of this and other condensed matter systems (with an eye towards
holography) geared towards a high energy theory audience, see e.g. the review \cite{McGreevy:2016myw}.}

From this perspective, it is natural to seek out additional examples of chiral
conformal field theories as potential generalizations of the fractional
quantum\ Hall effect to higher dimensions. Along these lines, there has
recently been significant progress in understanding the construction of
6D supersymmetric conformal field theories (6D\ SCFTs). An
important ingredient in these theories is that in the deformation away from
the conformal fixed point, there are always anti-chiral two-form potentials
with an anti-self-dual three-form field strength. This is a higher-dimensional
analog of a chiral boson, and as such, there is no known way to write a
Lorentz covariant action.

\begin{figure}[t!]
\centering
\includegraphics[trim={4cm 5cm 4cm 3cm},clip, height= 6cm]{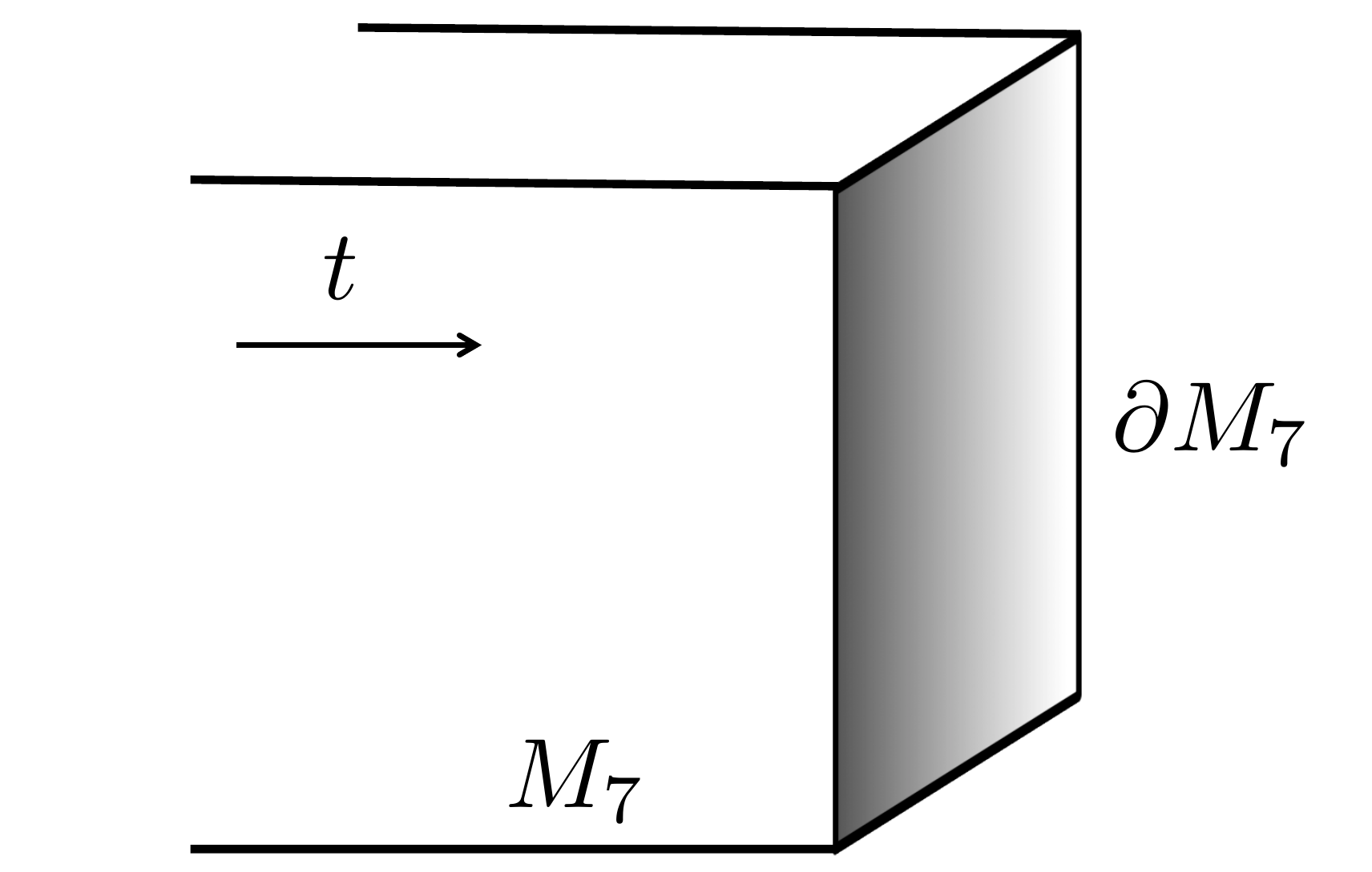}
\caption{A representation of the bulk-boundary geometry considered in this
paper. We denote the 7D bulk by $M_{7}$, the 6D boundary by $\partial M_{7}$
and time running into the boundary by $t$.}%
\label{Fig1}%
\end{figure}

Our generalization of the fractional quantum Hall effect will involve a 6D anti-chiral
theory of edge modes coupled to a $6+1$-dimensional bulk $M_{7}$ with Euclidean boundary
$\partial M_{7} = M_{6}$. The 7D action we propose to study has been considered before
in the high energy theory literature, both in the context of the topological sector of the
AdS/CFT correspondence \cite{Witten:1998wy, Aharony:1998qu, Maldacena:2001ss},
and also as an interesting topological field theory in its own right
\cite{Verlinde:1995mz, Belov:2006jd, Belov:2006xj, Freed:2012bs,
DelZotto:2015isa, Monnier:2017klz}:
\begin{equation}
S_{7D}=\frac{\Omega_{IJ}}{4\pi i}\underset{M_{7}}{\int}c^{I}\wedge dc^{J},
\label{7Daction}%
\end{equation}
with $c^{I}$ a collection of three-form potentials with gauge redundancy
$c^{I}\rightarrow c^{I}+db^{I}$ and subject to the imaginary-self-dual boundary condition (see
e.g. \cite{Belov:2006jd}):%
\begin{equation}
\ast_{6D}c^{I}|_{\partial M_{7}}=+ic^{I}|_{\partial M_{7}}\,.
\end{equation}
Here, our sign convention is such that if we analytically continue to
Lorentzian signature, the resulting boundary condition will yield an
anti-self-dual three-form, in accord with the convention adopted for 6D\ SCFTs
in references \cite{Heckman:2013pva, DelZotto:2014hpa, Heckman:2015bfa}. The
pairing $\Omega_{IJ}$ is an integral, positive definite symmetric matrix which
is the analog of the \textquotedblleft$K$-matrix\textquotedblright\ in the
context of the standard fractional quantum Hall effect.
The resulting 7D theory describes the low energy dynamics of $2+1$-dimensional
membranes. In the boundary, these will appear as 2D Euclidean strings coupled to
anti-chiral two-form potentials $b^{I}$ such that:
\begin{equation}
c^{I}=db^{I}.
\end{equation}
In this context, the matrix $\Omega_{IJ}$ governs the braiding statistics for
Euclidean strings in the boundary 6D system. The existence of this bulk action is in
some sense necessary to properly quantize the 6D theory
since it is otherwise impossible to simultaneously impose a self-duality
condition and quantization condition for three-form fluxes
\cite{Belov:2006jd}.\footnote{The reason for this is that there is no integral basis
of imaginary-self-dual three-forms on a generic six-manifold with compact three-cycles.
This is simply because the imaginary-self-duality condition depends on a choice of metric, and as such,
continuous variation of the metric destroys the chance to have an integral basis of imaginary-self-dual three-forms.
In the physical context, there is an additional caveat to this condition in the special case where $\det \Omega = 1$ since
then we have an invertible quantum field theory in the sense of reference \cite{Freed:2012bs}. In such
situations one expects to be able to decouple the bulk and boundary theories since the
boundary theory has a well-defined partition function.}

To realize both the integer and fractional quantum Hall effect, we activate a
background magnetic four-form field strength, namely one in which all four
legs thread the 6D boundary. Using the 7D perspective, we argue
that the many body wavefunction is constructed from correlation functions of
non-local operators of the form:%
\begin{equation}
\Phi(\Sigma)=\exp\left(  im_{I}\underset{\Sigma}{\int}b^{I}\right)  ,
\end{equation}
which from a 7D perspective involves replacing the integral of $b^{I}$ over a
Riemann surface by an integral of $c^{I}$ over a three-chain with boundary
$\Sigma$. Much as in the 2D case, our many body wavefunction factorizes as
the product of a piece controlled by correlation functions involving the
$\Phi$'s, and a Landau wavefunction which dictates the overall size of
droplets. Perhaps not surprisingly, this latter wavefunction is in turn
controlled by the background four-form flux.

Though this correlation function is likely to be quite
difficult to compute for an interacting 6D CFT, we find that in the case of a free theory
of anti-chiral two-forms, the evaluation is relatively straightforward, and can be
derived from general properties of conformal invariance. Said differently, the
absence of a Lagrangian formulation for 6D\ chiral CFTs does not impede our
analysis. The end result is somewhat more involved than that of the
2D Laughlin wavefunction, but even so, we find that it reduces to a quite similar form
in a zero slope limit where the membranes are large and rigid. Other limits
dictated by the relative energy scales set by the field strength and membrane tension
lead to deviations from this simple behavior.

In the context of 6D\ CFTs with supersymmetry (which can be realized in a geometric phase of F-theory),
the full list of $\Omega_{IJ}$ has actually been classified
\cite{Heckman:2013pva, DelZotto:2014hpa, Heckman:2015bfa,
Bhardwaj:2015xxa}. From a geometric perspective, these matrices are nothing
but the intersection pairings obtained from the resolution of orbifold
singularities of the form $%
\mathbb{C}
^{2}/\Gamma_{U(2)}$ for $\Gamma_{U(2)}$ a particular choice of discrete subgroup of $U(2)$.
The structure of this singularity is in turn controlled by continued
fractions \cite{jung, MR0062842, Riemen:dvq}:
\begin{equation}
\frac{p}{q}=x_{1}-\frac{1}{x_{2}-\frac{1}{x_{3}-...}}\,.
\end{equation}
From the perspective of the fractional quantum Hall effect, these are
interpreted as filling fractions for excitations above the ground state
\cite{Laughlin:1983fy, Haldane:1983xm, Halperin:1983zz, Halperin:1984fn,
Jain:1989tx}.

The 7D starting point also provides a unifying
perspective on a variety of lower-dimensional bulk topological systems coupled
to dynamical edge modes. In these systems, the analog of the $K$-matrix is
dictated by a tensor product $\Omega_{(7D)}\otimes\Omega_{(\text{intersect)}}%
$, with $\Omega_{(\text{intersect)}}$ the intersection pairing on the
compactification manifold. Indeed, even our 7D system can be viewed as the
compactification of an 11D theory of five-forms placed on a background of the
form $M_{7}\times M_{4}$.

The higher-dimensional unification in terms of this 11D theory
suggests an irresistible further extension to twelve dimensions, especially in the context
of F-theory, the non-perturbative formulation of IIB string theory.
With this in mind, we present a speculative conjecture on what such a 12D theory ought to look
like, showing that many elements can indeed be realized,
albeit for a supersymmetric theory in $10+2$ dimensions. To avoid pathologies
with having two temporal directions (such as moving along closed timelike curves),
we demand from the start that our theory is purely topological
in the bulk, namely the only propagating degrees
of freedom are localized along a $9+1$-dimensional spacetime.

The rest of this paper is organized as follows. First, in section
\ref{sec:BULK}, we discuss in more detail the bulk 7D topological field theory
which will form the starting point for our analysis. Using this perspective,
we determine the associated fractional conductivities for membranes, and also
present a formal answer for the analog of the Laughlin wavefunction. Next, in
section \ref{sec:BDRY} we evaluate the Laughlin wavefunction for a free
theory of anti-chiral tensors. In section \ref{sec:11D} we discuss the
spectrum of quasi-excitations, and its interpretation in terms of an
11D topological field theory of five-forms. In section \ref{sec:CPCT} we
briefly discuss compactifications of the 7D (and 11D) bulk topological field
theory to lower dimensions. Section \ref{sec:EMBED}\ discusses the embedding of the 7D theory in
M-theory, and section \ref{sec:Ftheory}\ presents our conjecture
on F-theory as a $10+2$-dimensional topological theory. Sections
\ref{sec:EMBED}\ and \ref{sec:Ftheory} are likely to be of more interest to a
high energy theory audience, and have been written so that they can be read
independently of the other sections. We present our conclusions in section
\ref{sec:CONC}.

\section{7D\ Bulk \label{sec:BULK}}

As mentioned in the introduction, our interest is in developing a
higher-dimensional generalization of the fractional quantum Hall system. In
this section, we lay out the main ingredients we use, focusing here on the 7D
bulk description. Indeed, because we at present lack a microscopic description
of 6D CFTs, it seems most fruitful to first develop the candidate effective
field theory which would govern the low energy physics.

Our starting point is the 7D action:\footnote{Following \cite{Maldacena:2001ss}%
, in order to properly define (\ref{7DCSagain}), we need to specify the global
nature of the fields $c$'s and their gauge transformations. In this
paper we assume that $c^{I} \sim c^{I} + Z^{p}_{2\pi\mathbb{Z}}(M_{7})$, where
$Z^{p}_{2\pi\mathbb{Z}}(M_{7})$ is the space of closed $p$-forms with integral periods on $M_{7}$. To properly
define this theory, it is helpful (much as in the ordinary Chern-Simons case) to view the 7-manifold as the boundary of an 8-manifold.
More formally, one needs the analog of spin structure on a three-manifold, which is played in seven dimensions
by a $4$-form Wu structure. This is necessary to have a properly defined theory of self-dual forms on the boundary. For
further discussion on this and related points, see references \cite{Belov:2006jd, Monnier:2016jlo}.
We shall neglect these subtleties in what follows since
much of our analysis will focus on coarse features of the 6D fractional quantum Hall effect which can all be recast in
terms of boundary correlators.}
\begin{equation}
\label{7DCSagain}S_{7D}[c^{I}]=\frac{\Omega_{IJ}}{4\pi i}\underset{M_{7}%
}{\int}c^{I}\wedge dc^{J},
\end{equation}
in which we further assume that the seven-manifold is given by a product of
the form:%
\begin{equation}
M_{7}=\mathbb{R}_{\text{time}}\times M_{6}\,.
\end{equation}
Said differently, when we proceed to quantize the theory, a wavefunction will
involve coordinates defined on $M_{6}$, which we then evolve from one time
slice to another.

There are various subtleties in quantizing such Chern-Simons-like theories,
and we refer the interested reader to the careful discussion presented in
references \cite{Belov:2006jd, Belov:2006xj, Belov:2005ze, Kapustin:2010hk}.
Following the discussion in reference \cite{Witten:1998wy}, we see that in a
gauge where we set:%
\begin{equation}
\text{Coulomb Gauge:\ }c_{tab}=0\,\text{,}%
\end{equation}
the canonical commutation relation for three-forms with all legs in spatial directions reads:
\begin{equation}
\lbrack c_{abc}^{I}(x),c_{def}^{J}(y)] = - 2\pi i\,\varepsilon_{abcdef}\left(
\Omega^{-1}\right)  ^{IJ}\delta^{6}(x-y)\,, \label{commrelate}%
\end{equation}
where $x$ and $y$ are coordinates on $M_{6}$. Note that to properly quantize
the theory, we therefore need to specify a Lagrangian splitting of the phase
space \cite{Belov:2006jd}.

Much as in the case of 3D Chern-Simons theory, there is a class of
observables of the form:%
\begin{equation}
\Phi_{m}\left(  S\right)  =\exp\left(  im_{I}\underset{S}{\int}c^{I}\right)  ,
\end{equation}
for $S\in H_{3}(M_{6},%
\mathbb{Z}
)$, and $m_{I}$ a vector of charges. The pairing $\Omega_{IJ}$ defines an
integral lattice $\Lambda$, and the $m_{I}$ take values in its dual
$\Lambda^{\ast}$.

The bulk correlation function of such observables is:
\begin{equation}
\langle\Phi_{m_{1}}(S_{1})\Phi_{m_{2}}(S_{2}) \dots\Phi_{m_{N}}(S_{N}%
)\rangle_{7D} = \exp\left[  2 \pi i \left(  m_{I}\left(  \Omega^{-1}\right)
^{IJ}m_{J}\right)  \sum_{1\leq i < j \leq N}L(S_{i},S_{j})\right]  \,,
\end{equation}
where $L(S_{i},S_{j})$ is the integral linking number of $S_{i}$ and $S_{j}$.
Given two bulk operators $\Phi_{m}\left(  S\right)  $ and $\Phi_{n}\left(
T\right)  $ for $m_{I} , n_{J} \in\Lambda^{\ast}$ and $S,T\in H_{3}(M_{6},%
\mathbb{Z}
)$, we get the braid relations \cite{Witten:1998wy}:%
\begin{equation}
\Phi_{m}\left(  S\right)  \Phi_{n}\left(  T\right)  =\Phi_{n}\left(  T\right)
\Phi_{m}\left(  S\right)  \times\exp\left(  2\pi i\left(  m_{I}\left(
\Omega^{-1}\right)  ^{IJ}n_{J}\right)  \left(  S\cdot T\right)  \right)  ,
\label{braiding}%
\end{equation}
where $S\cdot T$ is the intersection pairing for three-cycles in $M_{6}$.

The braiding relation of equation (\ref{braiding}) tells us that\ in the
quantum theory, we cannot simultaneously specify the periods of the three-form
for all three-cycles in $M_{6}$. Rather, we must take a maximal sublattice of
commuting periods and use this to specify the ground state(s). Calling this
lattice of three-cycles $L\subset H_{3}(M_{6},%
\mathbb{Z}
)$, the degeneracy of the ground state is:\footnote{A more formal way to
understand this relation is to note that states of the 7D bulk theory must
assemble into representations of the Heisenberg group \underline{$H$}%
$^{3}(M_{6},\Lambda^{\ast}/\Lambda)$. The maximal set of commuting fluxes
$L\subset$\underline{$H$}$^{3}(M_{6},\Lambda^{\ast}/\Lambda)$ leads to a
ground state degeneracy of dimension $\left\vert \Lambda^{\ast
}/\Lambda\right\vert ^{\dim L}$. Note also that in the case of a 3D theory on
$\mathbb{R}\times\Sigma$ for $\Sigma$ a genus $g$ Riemann surface, the
degeneracy would be $\left(  \det\Omega\right)  ^{g}$. Here we have neglected subtleties having
to due with the presence of torsion. Some additional comments on the effects of this (which do not alter
the degeneracy of the ground state) can be found in reference \cite{Monnier:2017klz}.}%
\begin{equation}
d_{\text{GND}}=\left(  \det\Omega\right)  ^{\dim L}.
\end{equation}

The three-form potential couples to $2+1$-dimensional membranes via integration
over its worldvolume. These degrees of freedom are
the analog of the electrons present in the fractional quantum Hall effect. In
contrast to that case, however, there is no guarantee that these membranes are
the genuine microscopic degrees of freedom for the 7D
system.\footnote{Moreover, there is a problem with formulating a theory of
first quantized membranes as fundamental degrees of freedom as this generically introduces
instabilities \cite{deWit:1988xki} (see also the review \cite{Taylor:1999qk}).
Here, however, there is not much of an issue since we are adopting an effective
field theory perspective. Indeed, even if the membrane disintegrates, the
macroscopic behavior we are considering should remain sensible since we are
only interested in collective features.} Indeed, from the perspective of
M-theory, it is widely expected that M2-branes are also simply a collective excitation.

Suppose now that we consider our bulk theory on the spacetime%
\begin{equation}
M_{7}=\mathbb{R}_{\leq0}\times M_{6}\,,
\end{equation}
so that at $t=0$, we realize the boundary $\partial M_{7}=M_{6}$, namely we
evolve a state from $t=-\infty$ to $t=0$. See figure \ref{Fig1} for a
depiction of this geometry. In this case, we need to impose suitable boundary
conditions for our system. Since we are assuming a Euclidean signature
boundary manifold, we take:%
\begin{equation}
\label{Euclideanbdc}\ast_{6D}c^{I}|_{\partial M_{7}}=+ic^{I}|_{\partial M_{7}%
}\,,
\end{equation}
which if we analytically continue to $M_{6}$ a Lorentzian signature manifold
(with mostly $+$'$s$ in the metric) would define an anti-self-duality relation.

From the perspective of the 6D boundary, we interpret the $c^{I}$'s as
three-form field strengths for two-forms $b^{I}$:%
\begin{equation}
c^{I}=db^{I}.
\end{equation}
The two-form couples to a Euclidean effective string, namely the
spatial section of our $2+1$-dimensional membrane. Including a membrane of
charge $m_{I}$ at a particular time can therefore be described by inserting
into the path integral the operator:%
\begin{equation}
\Phi_{m}(\Gamma)=\exp\left(  im_{I}\underset{\Gamma}{\int}c^{I}\right)  ,
\label{chains}%
\end{equation}
where $\Gamma=\mathbb{R}_{\leq0}\times\Sigma$ is a three-chain with boundary
$\Sigma$ a Riemann surface wrapped by the membrane.
\begin{figure}[t!]
\centering
\includegraphics[trim={4cm 5cm 4cm 3cm},clip, height= 6cm]{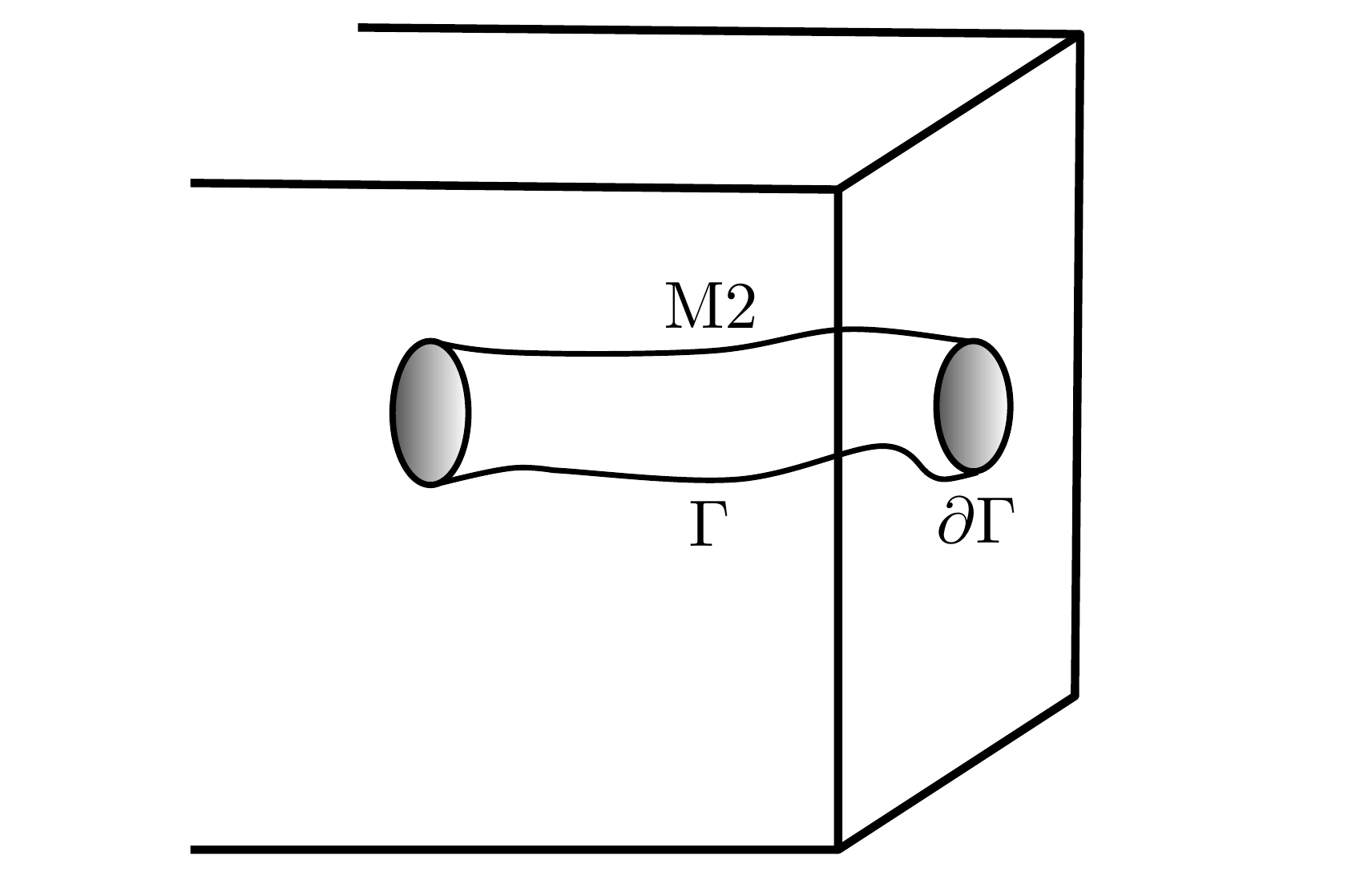}
\caption{State insertion of a membrane wrapping a three-cycle $\Gamma$ with
boundary $\partial\Gamma$.}%
\label{Fig2}%
\end{figure}

The choice of charge vector $m_{I}$ for the membrane depends on whether we
view it as a dynamic or static object in the 6D\ boundary theory. Indeed,
though the $m_{I}$ take values in the dual lattice $\Lambda^{\ast}$, we can
also entertain a special class of charges in the sublattice $\Lambda$ by
instead working with $m_{I}=\Omega_{IJ}n^{J}$ for $n^{J}\in\Lambda$. This
distinction will prove important when we come to the structure of the Laughlin
wavefunction where we will need to further restrict our choice of charges in
this way.

A background collection of membranes is conveniently described by adding a
source term $j_{I}$ to the action as follows:%
\begin{equation}
S_{7D}[c^{I},j]=\frac{\Omega_{IJ}}{4\pi i}\underset{M_{7}}{\int}c^{I}\wedge
dc^{J}-i\underset{M_{7}}{\int}j_{I}\wedge c^{I}.
\end{equation}
The equations of motion in the presence of a background source are:%
\begin{equation}
j_{I}=\frac{\Omega_{IJ}}{2\pi}dc^{J}.
\end{equation}
Adhering to the standard condensed matter terminology we dualize $j_{I}$ to a
three-form. The two-point function for these three-index objects defines a
six-index conductivity $\sigma_{abcdef}$, with all indices different.

Suppose now that we have a background three-form $C$ which couples to the
membranes. In this interpretation, the $c^{I}$ simply correspond to various
emergent gauge fields at low energies. The coupling between the two is:
\begin{equation}
S_{7D}[c^{I},C]=\frac{\Omega_{IJ}}{4\pi i}\underset{M_{7}}{\int}c^{I}\wedge
dc^{J}+\frac{\nu_{I}}{2\pi i}\underset{M_{7}}{\int}C\wedge dc^{I}.
\label{7Dbkgnd}%
\end{equation}
Physically, the membranes are all charged under $C$, which has field strength
$G$. In our conventions, the $G$-flux is quantized in units
of $2\pi$:\footnote{Strictly speaking, what is really required is that the
integration over a difference of two four-cycles is integral. There can
still be a half-integral shift, and this plays an important role in
compactifications of M-theory \cite{Witten:1996md}.}%
\begin{equation}
\frac{1}{2\pi}\underset{\text{4-cycle}}{\int}G\in%
\mathbb{Z}
.
\end{equation}
Much as in the lower-dimensional setting, we then obtain a fractional
conductivity given by:%
\begin{equation}
\sigma_{abcdef}= \frac{1}{2 \pi} \left(  \Omega^{-1}\right)  ^{IJ}\nu_{I}\nu_{J}\text{.}%
\end{equation}

Our discussion so far has focused on the bulk gapped system. Of course, it is
also important to understand the structure of the boundary theory. We expect
the many body wavefunction given by inserting a large number of
membrane states in the presence of a background four-form flux
to be in the same universality class as that of the genuine ground state.
Each membrane corresponds to the insertion of an operator of the form given by line
(\ref{chains}), for some choice of three-chain $\Gamma$ with boundary a
Riemann surface $\Sigma$, and some choice of charge vector $m_{I}$ in the dual
lattice of charges $\Lambda^{\ast}$. We label each such insertion in the path
integral by an operator $\Phi^{(i)}$ for $i=,1...,N$.\ The operator associated with a
background four-form flux follows from line
(\ref{7Dbkgnd}):%
\begin{equation}
\Phi_{\text{bkgnd}}=\exp\left(  \frac{\nu_{I}}{2\pi i}\underset{\mathbb{D}_{7}%
}{\int} c^{I}\wedge G\right) = \exp\left(  \frac{\nu_{I}}{2\pi i}\underset{\mathbb{D}_{6}%
}{\int} b^{I}\wedge G\right) ,
\end{equation}
where $\mathbb{D}_{7}=\mathbb{R}_{\leq0}\times\mathbb{D}_{6}$ is a
seven-dimensional domain, which at $t=0$ is given by a six-dimensional domain
$\mathbb{D}_{6}$ which has no overlap with the locations of the Riemann
surfaces $\Sigma_{(i)}$ wrapped by the membranes.

Putting all of this together, we expect the (unnormalized) many body
wavefunction for $N$ such membranes to take the form:%
\begin{equation}
\Psi_{\text{Laughlin}}=\left\langle \Phi^{(1)}...\Phi^{(N)}\Phi_{\text{bkgnd}%
}\right\rangle _{6D}, \label{Laughlin}%
\end{equation}
where the correlation function is evaluated in the boundary 6D theory.

Assuming that our 6D boundary theory is actually free, everything reduces
to the calculation of appropriate two-point functions:
\begin{equation}
\Psi_{\text{Laughlin}}=\underset{1\leq i<j\leq N}{\prod}\langle\Phi^{(i)}%
\Phi^{(j)}\rangle_{6D}\times\underset{1\leq i\leq N}{%
{\displaystyle\prod}
}\Psi_{\text{Landau}}^{(i)} \label{LaughlinAgain}%
\end{equation}
where we have introduced the unnormalized Landau wavefunction for a single
membrane moving in a background four-form flux:
\begin{equation}
\Psi_{\text{Landau}}^{(i)}=\langle\Phi_{\text{bkgnd}}\Phi^{(i)}\rangle_{6D}.
\end{equation}
Our goal will be to estimate $\Psi_{\text{Laughlin}}$ in various limits.

\section{Many Body Wavefunction \label{sec:BDRY}}

In the previous section we presented a bulk perspective on the 6D fractional
quantum Hall effect. Our aim in this section will be to extract additional
details on the structure of the ground state wavefunction. To obtain concrete
formulas, we focus on the case of a 6D CFT in flat space, namely
$\mathbb{R}^{6}$. Note that our result also generalizes to other conformally
flat spaces such as $S^{6}$, or a 6D ball (namely the hyperbolic space
$\mathbb{H}_{6}$). For ease of
exposition, we shall focus on the theory of a single emergent three-form
potential $c$, which couples to a uniform background magnetic four-form flux
$G=dC$ (locally), so we consider:
\begin{equation}
S_{7D}[c,C]=\frac{\Omega}{4\pi i}\underset{M_{7}}{\int}c\wedge dc+\frac
{1}{2\pi i}\underset{M_{7}}{\int}C\wedge dc.
\end{equation}
The generalization to the action of line (\ref{7Dbkgnd}) follows a similar
line of argument.

The starting point for our analysis is the observation that on the 6D
boundary, we have a theory of anti-chiral two-forms. In the case of a 2D chiral
boson in flat space, there is a well-known (non-covariant) action given in
reference \cite{Floreanini:1987as}. In principle one could
construct a similar action for the 6D anti-chiral two-form.

We shall not follow this route, but will instead simply appeal
to the fact that we have a 6D CFT, and use the resulting
structure of correlation functions for local operators. Indeed, we anticipate
that our analysis will generalize to more involved interacting theories.

With this in mind, we need to extract the two-point function for non-local
operators such as:
\begin{equation}
\left\langle \Phi_{m}(\Sigma) \Phi_{m^{\prime}}(\Sigma^{\prime}) \right\rangle
_{6D} = \left\langle \exp\left(  i m \underset{\Sigma}{\int} b \right)
\exp\left(  i m^{\prime} \underset{\Sigma^{\prime}}{\int} b^{\prime} \right)
\right\rangle _{6D},
\end{equation}
for some choice of charges $m$ and $m^{\prime}$, and Riemann surfaces $\Sigma$
and $\Sigma^{\prime}$ wrapped by the membranes. Here, the prime on $b^{\prime}$ serves
to remind us that the potential is supported on $\Sigma^{\prime}$. For now, we treat these
Riemann surfaces as fixed, though when we turn to the analysis of the Landau
wavefunction, we will show how to fix their mean field values. Since we are
dealing with a free field theory, we apply Wick's theorem to such correlation
functions to write:
\begin{equation}
\label{PhiPhiWick}\left\langle \Phi_{m}(\Sigma) \Phi_{m^{\prime}}%
(\Sigma^{\prime}) \right\rangle _{6D} = \exp\left\langle - m m^{\prime}
\underset{\Sigma}{\int} \underset{\Sigma^{\prime}}{\int} b b^{\prime}
\right\rangle _{6D}.
\end{equation}
We therefore need to extract the integrated two-point function for the
$b$-fields of our boundary theory. Strictly speaking, such a
correlation function is not gauge invariant. Note, however, that by
integrating over a closed Riemann surface, we should expect to obtain an
answer independent of a particular gauge. Said differently, our operators and
correlation functions are well-defined.

As we have already remarked, we do not have a covariant action for our anti-chiral
two-form. Indeed, this is not even an operator in the 6D CFT. Rather, we know
that the three-form field strength $h$ given locally by $db$ is a well-defined
operator. Our strategy will therefore be to compute the two-point
function for $h$, and to then integrate this two-point function over a pair of
three-chains with boundaries $\Sigma$ and $\Sigma^{\prime}$, respectively.

The answer in this case follows directly from that given for a theory of
two-forms with no self-duality constraint imposed. We follow the procedure
outlined in references \cite{AlvarezGaume:1983ig, Ganor:1998ve, Bastianelli:1999ab, Ganor:2000vt}.
Denoting the field strength for this non-chiral two-form by $h_{nc}$, the two-point function is:
\begin{equation}
\left\langle h_{a_{1}a_{2}a_{3}}^{(nc)}(x)h_{(nc)}^{b_{1}b_{2}b_{3}%
}(y)\right\rangle _{6D}=\frac{18}{\pi^{3}}\times\frac{1}{r^{6}}\left[
\delta_{\lbrack a_{1}}^{b_{1}}\delta_{a_{2}}^{b_{2}}\delta_{a_{3}]}^{b_{3}%
}-\frac{6r_{[a_{1}}r^{[b_{1}}\delta_{a_{2}}^{b_{2}}\delta_{a_{3}]}^{b_{3}]}%
}{r^{2}}\right]
\end{equation}
where we have introduced the relative separation:
\begin{equation}
r^{a}=x^{a}-y^{a}.
\end{equation}
Let us note that in this expression, there is in principle also a Dirac delta
function contact term. This involves details about the microscopic theory, and
can be removed by a suitable counterterm consistent with 6D conformal
invariance. We note that this expression holds both in Minkowski and Euclidean spacetimes
(by suitable choice of metric).

To reach the expression for the anti-chiral two-form theory, we apply a
projection to imaginary-self-dual field strengths, namely we set:
\begin{equation}
h=\frac{1}{2}(h_{(nc)}-i\ast_{6D}h_{(nc)}).
\end{equation}
The two-point function for the chiral theory is then:
\begin{equation}
\langle h_{a_{1}a_{2}a_{3}}(x)h^{b_{1}b_{2}b_{3}}(y)\rangle=\frac{9}{\pi^{3}%
}\frac{1}{r^{8}}\left(  r_{[a_{1}}r^{[b_{1}}\delta_{a_{2}}^{b_{2}}%
\delta_{a_{3}]}^{b_{3}]}+\frac{i}{6}r^{c}\varepsilon_{ca_{1}a_{2}a_{3}%
}^{\;\;\;\;\;\;\;\ \,[b_{1}b_{2}}r^{b_{3}]}\right)  ,
\end{equation}
which is a somewhat more involved expression than its counterpart in two dimensions.

To obtain the integrated two-point function for the chiral two-forms, we now
formally integrate this result over a pair of three-chains $\Gamma$ and
$\Gamma^{\prime}$ inside of $\mathbb{R}^{6}$:
\begin{equation}
\left\langle \Phi_{m}(\Sigma)\Phi_{m^{\prime}}(\Sigma^{\prime})\right\rangle
_{6D}=\exp\left\langle -mm^{\prime}\underset{\Gamma}{\int}\underset{\Gamma
^{\prime}}{\int}hh^{\prime}\right\rangle _{6D}. \label{formal}%
\end{equation}
where $\partial\Gamma=\Sigma$ and $\partial\Gamma^{\prime}=\Sigma^{\prime}$.
Returning to equation (\ref{LaughlinAgain}), we see that our expression for
the many body wavefunction:
\begin{equation}
\Psi_{\text{Laughlin}}=\underset{1\leq i<j\leq N}{\prod}\langle\Phi^{(i)}%
\Phi^{(j)}\rangle_{6D}\times\underset{1\leq i\leq N}{%
{\displaystyle\prod}
}\Psi_{\text{Landau}}^{(i)}%
\end{equation}
reduces to the calculation of these integrated two-point functions for the
three-form fluxes.

So far, we have followed a quite similar plan to what is typically done in the
2D fractional quantum Hall effect; We have expressed the Laughlin wave
function as a correlation function in a boundary CFT, and have also presented
a formal expression for its structure (see e.g. \cite{Moore:1991ks} as well as
the review in \cite{Tong:2016kpv}).

But in contrast to the case of the 2D system which involves point particles,
we are now faced with extended objects which carry an intrinsic tension:
\begin{equation}
T_{M2} = \frac{1}{(2 \pi\ell_{\ast})^{3}}.
\end{equation}
Depending on the strength of the four-form flux, the membrane may either puff
up to a large rigid object, or may instead be more accurately approximated by
a point particle. This will in turn affect the shapes of the Riemann surfaces
wrapped by the membranes.

We expect that a complete analysis will
involve a generalization of the loop equations in QCD to the case of
membranes, perhaps along the lines of references \cite{Ganor:1998ve, Ganor:1996nf}.
Even so, we can still use the structure of the correlation
functions just extracted to approximate these dynamics. Our goal in the
remainder of this section will be to characterize the typical size of a
membrane, and to then use this to extract the behavior of the Laughlin wave
function in various regimes. To this end, in the next subsection, we show that
the Landau wavefunction factor always leads to a certain amount of
``puffing up'' for the membrane in all six
spatial directions. This in turn depends on the particular profile for the
four-form flux. After this, we turn to two special limits. In the limit where
the membranes are very large compared to the intrinsic length scale
$\ell_{\ast}$, we show that the Laughlin wavefunction actually reduces to a
form quite close to that of the standard 2D Laughlin wavefunction. We also
consider the opposite regime of dilute four-form flux in which the membranes
are well-approximated by point particles.

\subsection{Landau Wavefunction}

Our aim in this subsection will be to extract additional details on the
structure of the Landau wavefunction factor appearing in equation
(\ref{Laughlin}), namely, the correlation function:
\begin{equation}
\Psi_{\text{Landau}}=\left\langle \exp\left(  \frac{\nu}{2 \pi i}\underset{\mathbb{D}_{6}%
}{\int}G\wedge b\right)  \exp\left(  im\underset{\Sigma}{\int}b\right)
\right\rangle _{6D}=\left\langle \Phi_{\text{bkgnd}}\exp\left(
im\underset{\Sigma}{\int}b\right)  \right\rangle _{6D}.
\end{equation}
One of the implicit assumptions we have made up to this point is that the
Riemann surface $\Sigma$ is held fixed. We now show that the presence of the
four-form flux actually causes the membrane to puff up. As mentioned at the
beginning of this section, we take the four-form flux to be uniform, and
write:
\begin{equation}
G=\frac{1}{4!}G_{abcd}dx^{a}\wedge dx^{b}\wedge dx^{c}\wedge dx^{d}%
\end{equation}
where $x^{a}$ are local coordinates on $\mathbb{R}^{6}$ and we have taken some
choice of constant $G_{abcd}$. Suppose that we work in the limit where the
Riemann surface $\Sigma$ is small. Then, we can parameterize its location, to
leading order, by the position of the center of mass, which we denote by
coordinates $y^{a}$. Applying the Laplacian  in the $y$
variable to the correlation function yields:
\begin{equation}
\Delta_{(y)}\left\langle \underset{\mathbb{D}_{6}}{\int}\frac{G}{2\pi}\wedge
b\underset{\Sigma}{\int}b^{\prime}\right\rangle _{6D}= - \underset{\mathbb{D}%
_{6}}{\int}\frac{G}{2\pi}\wedge \ast_{6D} \delta_{\Sigma}^{(4)}.
\label{DeltaAction}%
\end{equation}
To reach the righthand side, we have used the fact that the membrane is
--by definition-- localized along $\Sigma$, so we know that it has a delta
function support for its source. Observe that in our integral, the only legs
of $G$ which actually participate are those which are normal to the Riemann
surface $\Sigma$. We denote these local coordinates by $y_{\bot}$.

\begin{figure}[t!]
\centering
\includegraphics[trim={5cm 6cm 5cm 5cm},clip, height= 6cm]{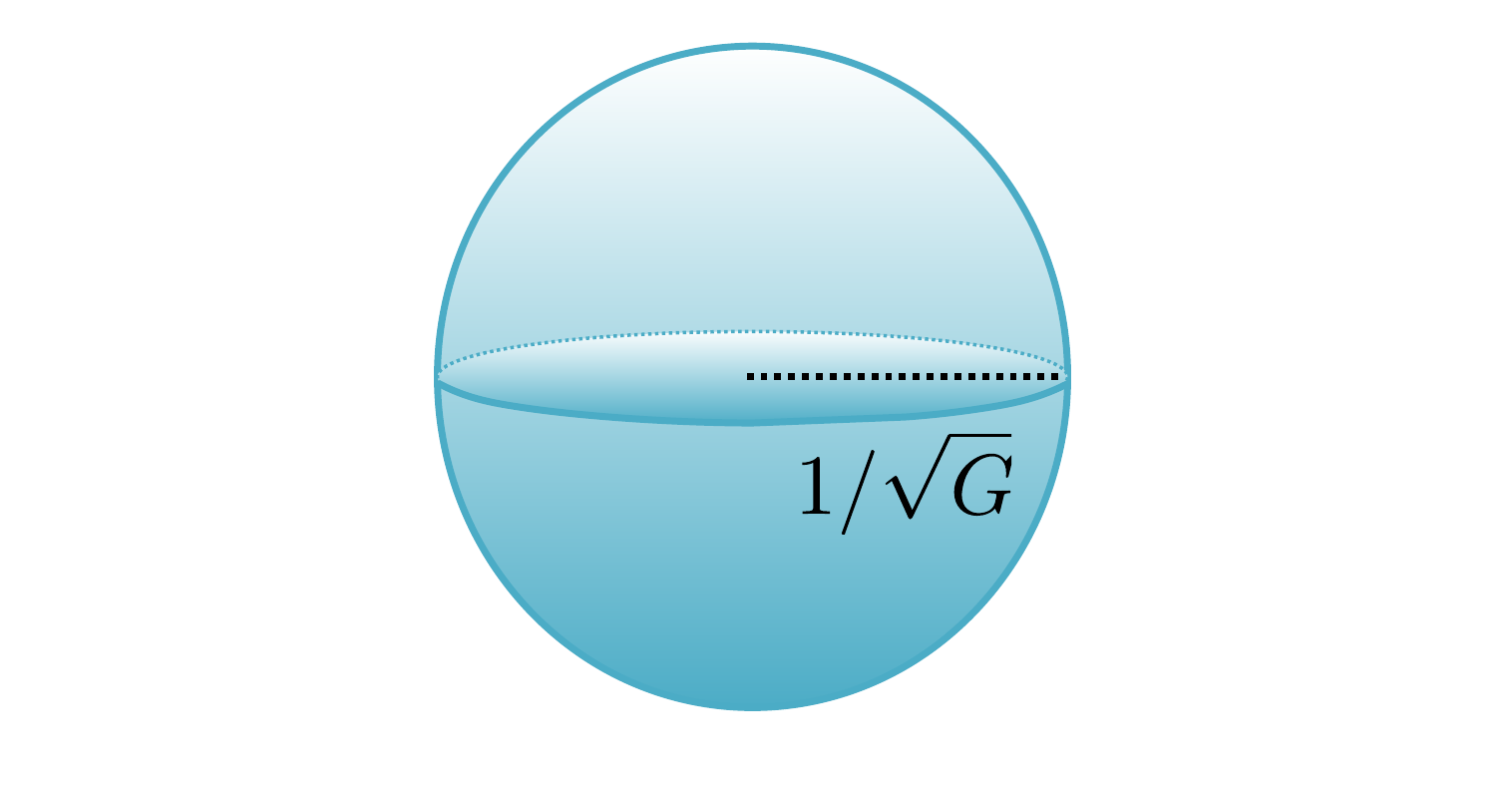}
\caption{Depiction of a quantum Hall droplet of characteristic radius $\sim 1/\sqrt{G}$.}%
\label{Fig3}%
\end{figure}

On the other hand, we also know that if we simply consider the action of the
Laplacian in these four directions, then we have:
\begin{equation}
\Delta_{\bot}y_{\bot}^{2}=8.
\end{equation}
Integrating equation (\ref{DeltaAction}), we conclude that our integral is, in
this limit, given by a quadratic form in the normal directions:
\begin{equation}
\left\langle \nu \underset{\mathbb{D}_{6}}{\int}\frac{G}{2\pi}\wedge
b\underset{\Sigma}{\int}b^{\prime}\right\rangle _{6D}= - \frac{y_{\bot}^{2}}%
{8}\rho_{\bot},
\end{equation}
where we have introduced the localized flux density:%
\begin{equation}
\rho_{\bot}= \nu \underset{\mathbb{D}_{6}}{\int} \frac{G}{2\pi}\wedge
\ast_{6D} \delta_{\Sigma}^{(4)}. \label{rholoc}%
\end{equation}

At this point, we can now see why the presence of a background four-form flux
prevents the membrane from collapsing to zero size. Indeed, from the
above analysis, we see that if we consider the width of the membrane in the
directions normal to $\Sigma$, there is a characteristic size $\ell_{\bot}$ indicating the
spread of the associated Gaussian:
\begin{equation}
\label{maglength}\frac{1}{2 \ell_{\bot}^{2}} \sim\frac{G_{\bot}}{16 \pi}%
\end{equation}
where $G_{\bot}$ denotes the component of the four-form flux with all legs
transverse to $\Sigma$. Continuing in the same vein, we see that provided we
have at least three independent contributions to the four-form flux, we always
get a puffed up membrane. Said differently, there is no direction we can place
a membrane so that it is \textit{not} polarized.

It is helpful at this point to compare with the standard case of the 2D
quantum Hall effect. There, we also have a droplet size, with characteristic
length set by $\ell\sim1/\sqrt{F}$, in appropriate units. Here, we see the
analog of this formula in equation (\ref{maglength}). In contrast to that
case, however, we see that since the four-form flux always picks a preferred
set of directions in the 6D space, we actually get a tensor of such
characteristic lengths.

\subsection{Zero Slope Limit of the Membrane}

In the limit of a large magnetic flux, we also expect the worldvolume theory
of the membrane to simplify. Here we present a brief sketch of how we expect
such a simplification to occur. Some elements of our discussion are
necessarily schematic, but we anticipate it will be useful for further investigations.

Consider the topological coupling between the three-form
potential and the membrane:
\begin{equation}
S_{\text{M2}}\supset\mu_{3}\int_{M_{3}}i^{\ast}C_{(3)},
\end{equation}
where $\mu_{3}$ is an overall constant of proportionality set by the tension
of the membrane, $i:M_{3}\rightarrow M_{7}$ is the embedding of the
worldvolume of the membrane in the 7D spacetime, and $i^{\ast}C_{(3)}$ denotes
the pullback of the bulk three-form onto the worldvolume. The embedding of the
membrane is captured by fields $X^{a}(\sigma^{1},\sigma^{2}%
,t)$. So, we can alternatively write the form of this coupling in components,
where we absorb various numerical pre-factors into the overall coupling:%
\begin{equation}
S_{M2}\supset\mu_{3}\int_{M_{3}}C_{abc}dX^{a}\wedge dX^{b}\wedge
dX^{c}.\label{couplar}%
\end{equation}
We are in particular interested in the special case of a large, uniform
magnetic G-flux. In this limit, we anticipate that the action is actually
dominated by this topological term. We refer to this as the zero-slope limit
action:%
\begin{equation}
S_{\text{zero-slope}}=\mu_{3}\int_{M_{3}}G_{abcd}X^{a}dX^{b}\wedge
dX^{c}\wedge dX^{d}.\label{zeroslope}%
\end{equation}
The general algebraic structure associated with $C$ flat and large
has been considered quite extensively in the literature (see e.g.
\cite{Berman:2007bv, Ho:2008nn, Ho:2008ve} and references therein).\ As far as
we are aware, however, the case of large G-flux has not been considered in
much detail. Part of the complication with this sort of coupling is that it
leads to a generalization of the Poisson bracket to a Nambu
2-bracket \cite{Nambu:1973qe}. The quantum theory contains many technical complications, including the
appearance of non-associative and non-commutative algebraic structures.

We bypass these complications (interesting though they may be) by focusing exclusively on the low energy
limit, i.e. leading derivative contributions to the effective theory. This
is accomplished most cleanly in the limit where we have a large background
G-flux, since in this case, the membranes are polarized, and most fluctuations
of the worldvolume will decouple.

Consider, for example, the expansion of the fields
$X^{a}(\sigma^{1},\sigma^{2},t)$. In the limit of slow fluctuations, we can
write:%
\begin{equation}
X^{a}(\sigma^{1},\sigma^{2},t)=X_{\text{cm}}^{a}(t)+P_{\text{cm}}^{a}%
(\sigma)+...,
\end{equation}
where to leading order, the $X_{\text{cm}}$'s have no position dependence, and
the $P_{\text{cm}}$'s have no time dependence. In this limit, then, we see
that in the action of equation (\ref{zeroslope}), each $X^{a}$ field can
support at most one worldvolume derivative. Consequently, we can integrate by
parts and present the action in the suggestive form:%
\begin{equation}
S_{\text{zero-slope}}=\mu_{3}\int_{M_{3}}G_{abcd}\varepsilon^{\mu\nu\rho
}\left(  X^{[a}\partial_{\mu}X^{b]}\right)  \partial_{\nu}\left(
X^{[c}\partial_{\rho}X^{d]}\right)  .
\end{equation}
Observe that the composite operators $X^{[a}\partial_{\mu}X^{b]}$ define a
collection of abelian gauge fields on the three-dimensional space:%
\begin{equation}
Y_{\mu}^{ab}=X^{[a}\partial_{\mu}X^{b]}.
\end{equation}
Interpreted in this way, we see that the G-flux can be split according to
pairs of indices $ab$ and $cd$, and with respect to these indices, it is a
symmetric matrix. Combining the parameter $\mu_3$ with that from $G$, we can now
present the canonical form of the zero slope limit in terms of the gauge
fields $Y_{\mu}^{ab}$ and a dimensionless matrix of couplings $K_{ab,cd}$:
\begin{equation}
S_{\text{zero-slope}}=\frac{K_{ab,cd}}{4\pi i}\int_{M_{3}}Y^{ab}\wedge
dY^{cd}.
\end{equation}
Consider now the quantization of this worldvolume theory.
In the gauge where $Y^{ab}_{0}=0$, the canonical commutation relation is:%
\begin{equation}
K_{ab,cd} [Y_{\mu}^{ab}(\sigma),Y_{\nu}^{cd}(\sigma^{\prime
})]= - 2 \pi i \varepsilon_{\mu\nu}\delta^{2}(\sigma-\sigma^{\prime}).
\end{equation}

The above commutation relation is the direct analog of what one finds for
electrons moving in a large magnetic field. Indeed, in that context, the resulting 1D topological
quantum mechanics is the starting point for the zero slope limit of open string theory in the presence
of a large Neveu-Schwarz B-field \cite{Seiberg:1999vs}. It is quite tempting
to also consider the limit of a large number of M2-branes, say in M-theory.
The backreaction of the four-form flux on an individual membrane should then
give, in a suitable limit, a Chern-Simons action. It would be interesting to
connect these observations to the constructions presented in
\cite{Aharony:2008ug, Aharony:2008gk}.

\subsection{Large Rigid Limit}

Now that we have determined the impact of the four-form flux on the size of
the droplets, we turn to the evaluation of the rest of the Laughlin wave
function in line (\ref{LaughlinAgain}). This requires us to specify a pair of
Riemann surfaces $\Sigma$ and $\Sigma^{\prime}$, as well as a pair of charges
$m$ and $m^{\prime}$. In the limit where the background magnetic flux is very
large, we see that the size of the membrane in the transverse directions is
quite small, going roughly as $\ell_{eff}\sim1/\sqrt{G}$. In this limit, then,
we approximate the membranes as wrapping very large Riemann surfaces (in units
of the membrane tension length), and very thin in the transverse directions.
To perform explicit computations, it is helpful to introduce a complex
structure for $\mathbb{R}^{6}$, writing $\mathbb{C}^{3}$ with local
coordinates $u,v,w$ such that the two Riemann surfaces are locally defined by
the equations:%
\begin{equation}
\Sigma=\left\{  u=u_{0}\right\}  \cap\left\{  w=0\right\}  \text{ \ \ and
\ \ }\Sigma^{\prime}=\left\{  u=u_{0}^{\prime}\right\}  \cap\left\{
v=0\right\}  \text{,}%
\end{equation}
so that the common normal coordinate for both Riemann surfaces is
parameterized by the $u$-plane, and the holomorphic coordinate:%
\begin{equation}
\xi=u_{0}-u_{0}^{\prime}%
\end{equation}
specifies the separation between the two Riemann surfaces.

To evaluate the correlation function of $\Phi(\Sigma)$ and $\Phi(\Sigma^{\prime})$
in this limit, we observe that in the two-point function for the $b$ field, we are
integrating (in momentum space) over the $w$ and $v$ directions. Consequently,
we are actually working in the limit of low momentum in these two directions.
The correlation function will therefore be dominated by momenta in the
$u$-plane. With this in mind, we see that our problem reduces to a
two-dimensional system.

In the related context where we compactify our 6D CFT on the space
$\mathbb{R}^{2}\times\Sigma\times\Sigma^{\prime}$, the low energy theory on
$\mathbb{R}^{2}$ is governed by a collection of chiral and anti-chiral bosons
which all descend from the anti-chiral two-form. Indeed, we can decompose the
$b$-field on shell as:%
\begin{equation}
b=\phi^{i}(\xi)\omega_{i}+\widetilde{\phi}\,\,\widetilde{^{i}}(\overline{\xi
})\omega_{\,\widetilde{i}},
\end{equation}
where $\omega_{i}$ is a basis of harmonic anti-self-dual two-forms on $\Sigma
\times\Sigma^{\prime}$, and $\omega_{\,\widetilde{i}}$ is a basis of
self-dual two-forms on $\Sigma\times\Sigma^{\prime}$. Here, the index
$i=1,...,b_{2}^{-}(\Sigma\times\Sigma^{\prime})$ and $\widetilde{i}%
=1,...,b_{2}^{+}(\Sigma\times\Sigma^{\prime})$ so that the $\phi^{i}(\xi)$ are
chiral bosons and the $\widetilde{\phi}\,\,\widetilde{^{i}}(\overline{\xi})$
are anti-chiral bosons. Integrating over the Riemann surfaces, we find that
the correlation function reduces to:%
\begin{equation}
\left\langle \exp\left(  im\underset{\Sigma}{\int}b\right)  \exp\left(
im^{\prime}\underset{\Sigma^{\prime}}{\int}b\right)  \right\rangle _{2D}%
=\xi^{\rho}\bar{\xi}^{\,\widetilde{\rho}}, \label{2dlimit}%
\end{equation}
where in general the values of the $\rho$ and $\widetilde{\rho}$ depend on
integrating the basis of self-dual and anti-self-dual forms over the Riemann
surfaces. This in turn depends on the details of the metric. There is,
however, an important aspect of this correlator which is protected by topology
(see e.g. \cite{Ganor:1996xg, Witten:1999vg, Seiberg:2011dr, Apruzzi:2016nfr}%
):%
\begin{equation}
\rho-\widetilde{\rho}=-(m\Omega^{-1}m^{\prime})(\Sigma\cdot\Sigma^{\prime}).
\end{equation}
Note that single-valuedness of the associated OPE requires us to work in terms
of charges $m$ and $m^{\prime}$ which scale in appropriate units of $\Omega$,
as per our discussion below equation (\ref{chains}). The simplest possibility
is to take $m=m^{\prime}=\Omega$, though more generally we can contemplate
$m=n\Omega$ and $m^{\prime}=n^{\prime}\Omega$ so that the difference becomes:%
\begin{equation}
\rho-\widetilde{\rho}=-(n\Omega n^{\prime})(\Sigma\cdot\Sigma^{\prime}).
\end{equation}

Returning to the evaluation of our correlation function, we see that there is
a rather close similarity to the case of line (\ref{2dlimit}). The main
difference is that in the limit we have just taken, we have discarded various
global data such as the topology of the Riemann surfaces. By construction,
however, we have assumed that the only intersections occur when $\xi=0$.
Consequently, we see that we can essentially carry over unchanged the
calculation in the 2D limit. The precise values of the exponents $\rho$ and $\widetilde{\rho}$ also
require information about the explicit choice of metric as well as
the dynamics of the membranes moving in a background charge density.

With this in place, we now generalize to
other configurations of affine planes. It is helpful to introduce holomorphic
homogeneous coordinates $Z^{\alpha}$ with $Z^{4}=1$.\footnote{A curious
feature of this formulation is the appearance of homogeneous
coordinates, and therefore a $\mathbb{CP}^{3}$. Additionally, by specifying
our Riemann surfaces by a pair of points in $\mathbb{CP}^{3}$, we also obtain the
standard correspondence between twistor space and the
complexification of conformally compactified four-dimensional Minkowski space \cite{Penrose:1967wn}.
The presence of a background four-form flux also suggests a non-commutative (possibly
covariant)\ deformation of this space (see e.g. \cite{Heckman:2011qt,
Heckman:2011qu, Heckman:2014xha}). It would be interesting to develop a
four-dimensional interpretation for the results of this paper.} The Riemann
surfaces can then be specified as:%
\begin{align}
\Sigma &  =\left\{  f_{\alpha}Z^{\alpha}=0\right\}  \cap\left\{  g_{\alpha
}Z^{\alpha}=0\right\} \\
\Sigma^{\prime}  &  =\left\{  f_{\alpha}^{\prime}Z^{\alpha}=0\right\}
\cap\left\{  g_{\alpha}^{\prime}Z^{\alpha}=0\right\}  .
\end{align}
In this case, the analog of the holomorphic separation between the two Riemann
surfaces is now given by:%
\begin{equation}
\xi_{\Sigma,\Sigma^{\prime}}=\varepsilon^{\alpha\beta\gamma\delta}f_{\alpha
}g_{\beta}f_{\gamma}^{\prime}g_{\delta}^{\prime},
\end{equation}
Assuming we remain in the rigid limit for all surfaces, we see that the resulting
contribution to the Laughlin wavefunction in line (\ref{LaughlinAgain}) takes
the form:%
\begin{equation}
\underset{1\leq i<j\leq N}{\prod}\langle\Phi^{(i)}\Phi^{(j)}\rangle_{6D}%
\sim\underset{1\leq i<j\leq N}{\prod}\xi_{ij}^{\rho_{ij}}\bar{\xi}%
_{ij}^{\,\widetilde{\rho}_{ij}},
\end{equation}
in the obvious notation.

\subsection{Point Particle Limit}

We can also evaluate the form of the many body wavefunction in the limit where
the relative separation between a pair of Riemann surfaces is quite large. In
this case, the membranes are well-approximated by point particles moving in
six spatial dimensions, and interacting via exchange of the bulk 7D
three-form.\ We can therefore proceed in two complementary ways. On the one
hand, we can simply calculate the scattering amplitude between two
non-relativistic M2-branes. Alternatively, we can work directly in terms of
the 6D\ CFT, and compute the long distance limit of the two-point function for
the anti-chiral two-forms, suitably integrated over the small Riemann surfaces.

\begin{figure}[t!]
\centering
\includegraphics[trim={4cm 5cm 4cm 3cm},clip, height= 6cm]{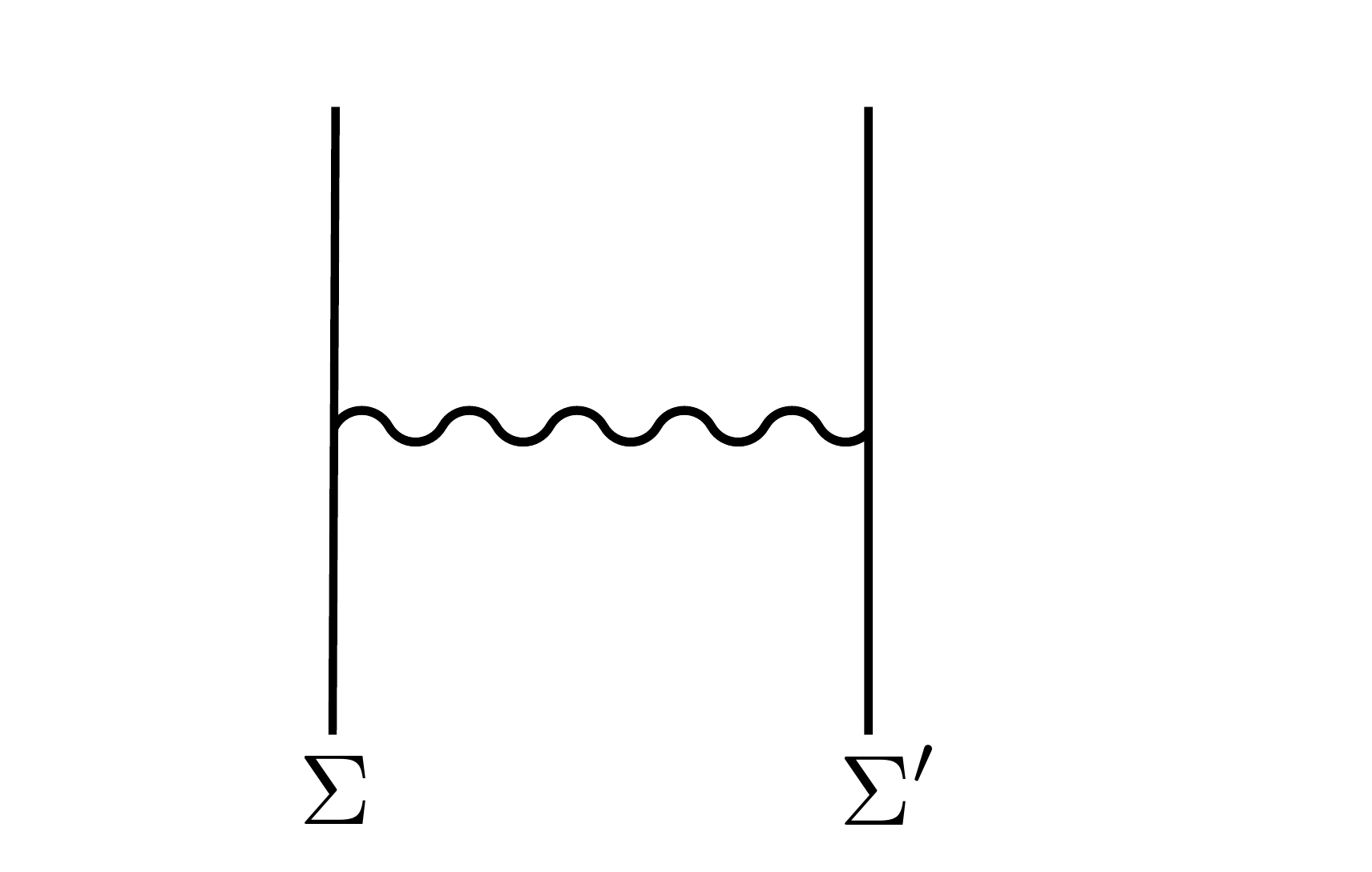}
\caption{Exchange of an anti-chiral two-form between two membranes wrapped on Riemann surfaces
$\Sigma$ and $\Sigma^{\prime}$.}%
\label{Fig4}%
\end{figure}

In either case, the problem reduces to that of a scattering amplitude. We
calculate the spin averaged value, neglecting issues of fine structure.
Indeed, if we were to treat the Riemann surface as fixed, we would get a four-form source given by the
corresponding delta function $\delta_{\Sigma}^{(4)}$, which dualizes to a
two-form $\varepsilon_{cd}$ and couples to the $b$-field via $\varepsilon
^{cd}b_{cd}$. At this point, it is convenient to work in Feynman gauge for the
two-point function of the non-chiral two-form:%
\begin{equation}
\langle b_{ab}^{\text{(nc)}}(x)b_{cd}^{\text{(nc)}}(y)\rangle=\frac{1}%
{2\pi^{3}}\frac{\eta_{ac}\eta_{bd}-\eta_{ad}\eta_{bc}}{(x-y)^{4}}\,,
\label{nc2form}%
\end{equation}
The spin averaged amplitude is then given by:%
\begin{equation}
\mathcal{A}_{\text{(nc)}}=-\left(  m_{(i)}\Omega^{-1}m_{(j)}\right)
\times\left(  \text{Vol}(\Sigma_{(i)})\times\text{Vol}(\Sigma_{(j)})\right)
\times\frac{1}{\pi^{3}}\frac{1}{(x_{(i)}-x_{(j)})^{4}}\,.
\end{equation}

Now, the amplitude receives two equal contributions, one from a basis of
chiral two-forms and another from anti-chiral two-forms, with no cross-terms between the two.
Because of this, the chiral case is half as large. Putting this
together, we reach our estimate for the correlation function in this limit:%
\begin{equation}
\langle\Phi^{(i)}\Phi^{(j)}\rangle_{6D}=\exp\left(  \left(  m_{(i)}\Omega
^{-1}m_{(j)}\right)  \times\left(  \text{Vol}(\Sigma_{(i)})\times
\text{Vol}(\Sigma_{(j)})\right)  \times\frac{1}{2\pi^{3}}\frac{1}%
{(x_{(i)}-x_{(j)})^{4}}\right)  .
\end{equation}
Observe that as we separate the particles, this correlator tends to one. In
the opposite limit, the apparent divergence is cut off by the short distance
behavior already described in the previous subsection. The crossover between
these two regimes occurs precisely when the characteristic size of the Riemann
surface becomes comparable to the separation.

\section{Quasi-Branes and Quasi-Dranes \label{sec:11D}}

As mentioned at the beginning of section \ref{sec:BDRY}, we can extend
this analysis to more general pairings $\Omega_{IJ}$. Indeed,
there is a well-known interpretation of this in the standard fractional
quantum Hall effect in terms of quasi-particle excitations / holes and their
associated emergent gauge fields. To better understand this in our system, it
is actually helpful to view the 7D bulk topological field theory as obtained
from the compactification of an 11D theory of a single abelian five-form with action:
\begin{equation}
S_{11D}=\frac{1}{4\pi i}\underset{M_{11}}{\int}C_{(5)}\wedge dC_{(5)}.
\end{equation}
Compactifying on a four-manifold, we assume the 11D spacetime takes the form
of a Cartesian product $M_{11}=M_{7}\times M_{4}$. There is a decoupled sector
given by a theory of three-forms. To study the structure of this subsystem, we
can consider a basis of two-cycles $\Sigma_{I}$ with pairing $\Omega
_{IJ}=-\Sigma_{I}\cap\Sigma_{J}$. Dual to these cycles are harmonic two-forms
$\omega^{I}$. To perform the reduction and maintain an integral basis of
fields, it is actually most convenient to work in terms of the related basis
of two-forms $\omega_{I}=\Omega_{IJ}\omega^{J}$. Note that we also have:%
\begin{equation}
\Omega_{IJ}=-\underset{M_{4}}{\int}\omega_{I}\wedge\omega_{J}.
\end{equation}
Expanding the five-form in terms of a basis of harmonic two-forms $\omega_{I}$
on $M_{4}$ yields:%
\begin{equation}
C_{(5)}=c_{(3)}^{I}\wedge\omega_{I}+...,
\end{equation}
where the other terms in \textquotedblleft...\textquotedblright\ refer to
decoupled sectors. In this 11D theory, the degrees of freedom in
the boundary are Euclidean D3-branes. These are wrapped over two-cycles of
$M_{4}$, and this gives rise to the membranes discussed in the previous section.
In this geometric construction, we take D3-branes wrapped over
collapsing two-cycles. This leads to effective strings in six
dimensions, which at the conformal fixed point have vanishing tension.

Indeed, in F-theory, the construction of 6D\ SCFTs involves compactification on
a singular base $\mathcal{B}=\mathbb{C}^{2}/\Gamma_{U(2)}$ with $\Gamma
_{U(2)}$ a discrete subgroup of $U(2)$. Not all discrete subgroups realize a
6D SCFT, in part because they are incompatible with the existence of an
elliptically fibered Calabi-Yau threefold with base $\mathcal{B}$. In the
resolved phase, we have a generalization of Dynkin diagrams of ADE\ type. In
fact, only the A- and D- series can have curves of self-intersection different
from $-2$. For additional details on the construction of 6D SCFTs in
F-theory, see references \cite{Heckman:2013pva, Heckman:2015bfa} as well as \cite{Morrison:2012np}.

An intriguing feature of 6D SCFTs with an F-theory origin is the appearance of continued
fractions such as:%
\begin{equation}
\frac{p}{q}=x_{1}-\frac{1}{x_{2}-\frac{1}{x_{3}-...}}. \label{pq}%
\end{equation}
As discussed in the introduction, the fractional quantum Hall effect also exhibits a
sequence of continued fractions, and these numbers specify filling fractions for the spectrum of physical
excitations above the ground state \cite{Laughlin:1983fy, Haldane:1983xm,
Halperin:1983zz, Halperin:1984fn, Jain:1989tx}. From an 11D perspective we can
explain the appearance of this structure in terms of geometrical properties of
the internal directions.

For example, in the generalization of an A-type base, this is a collection of
curves of self-intersection $-x_{i}$ with pairing:%
\begin{equation}
\Omega=\left[
\begin{array}
[c]{ccccc}%
x_{1} & - 1 &  &  & \\
- 1 & x_{2} & - 1 &  & \\
& - 1 & ... & - 1 & \\
&  & - 1 & x_{k-1} & 1\\
&  &  & - 1 & x_{k}%
\end{array}
\right]  .
\end{equation}
The orbifold singularity is then given by the group action on $\mathbb{C}^{2}$
with local coordinates $u$ and $v$ as:%
\begin{equation}
(u,v)\mapsto\left(  \zeta u,\zeta^{q}v\right)  ,
\end{equation}
for $\zeta$ a primitive $p^{th}$ root of unity (for example $\zeta=\exp(2\pi
i/p)$).

Here we see that the spectrum of \textquotedblleft
quasi-particles\textquotedblright\ are actually Euclidean effective strings dictated by
the continued fraction of line (\ref{pq})! Moreover, we also know that at
least for 6D SCFTs realized in F-theory, the space of possible
pairings $\Omega_{IJ}$ is tightly constrained. For example, the
self-intersection numbers must always obey $1 \leq x_{i} \leq 12$. Additionally,
further blowups of the base do not shift the value of $p$ or $q$, but do
introduce additional quasi-branes \cite{DelZotto:2015isa, Apruzzi:2017iqe}.

Turning the discussion around, it is natural to ask whether there is a top
down interpretation of \textquotedblleft quasi-dranes\textquotedblright\ (the
negation of a brane). Physically, this would appear to descend from
anti-Euclidean D3-branes. We leave an analysis of this issue for future work.

\section{Compactification \label{sec:CPCT}}

One of the general paradigms of many condensed matter systems is the presence
of a gapped bulk coupled to edge modes. It is also widely
believed that this gapped phase is described by a topological field theory.
Motivated by these considerations, in this section we use our
higher-dimensional starting point as a tool in generating consistent examples
of such phenomena. Our aim in this section will be to understand the class of
theories generated from compactifications of our 7D bulk theory, and the
corresponding edge modes. For simplicity, we focus on the case where there bulk flux is
switched off.

In fact, following up on our discussion in section \ref{sec:11D}, it is
helpful to actually begin with an 11D bulk theory of five-forms with action:%
\begin{equation}
S_{11D}=\frac{1}{4\pi i}\underset{M_{11}}{\int}C_{(5)}\wedge dC_{(5)}.
\end{equation}
Some aspects of this theory have been studied in \cite{Belov:2006jd,
Belov:2006xj}. Suppose now that we restrict the form of the 11D spacetime to
be a product of the form:%
\begin{equation}
M_{11}=M_{11-p}\times M_{p},
\end{equation}
in which $M_{11-p}$ is a Lorentzian signature spacetime and $M_{p}$ is
a Euclidean signature space. From the perspective of the boundary theory, we can take
a limit in the space of metrics where $M_{p}$ is relatively small compared
with $\partial M_{11-p}$. In this sense, we can \textquotedblleft
compactify\textquotedblright\ and reach a lower-dimensional theory defined
solely on $M_{11-p}$.

To better understand the resulting theory, we decompose $C_{(5)}$ into a basis
of harmonic differential forms defined on $M_{p}$:%
\begin{equation}
C_{(5)}=\underset{i=0}{\overset{p}{\sum}}\underset{k_{i}%
=1}{\overset{b_{\text{cpct}}^{i}(M_{p})}{\sum}}C_{k_{i}}^{(5-i)}\wedge
\omega_{k_{i}}^{(i)},
\end{equation}
namely, we sum over all compact harmonic $i-$forms with degeneracy label
$k_{i}$, and also sum over all choices of $i$. Here, $b_{\text{cpct}}%
^{i}(M_{p})=\dim H_{\text{DR,cpct}}^{i}(M_{p})$.\footnote{Note that since we
do not assume $M_{p}$ is compact, we cannot assert a relation
between $b_{\text{cpct}}^{i}(M_{p})$ and $b_{\text{cpct}}^{p-i}(M_{p})$.}
There is a canonical pairing on $M_{p}$ between an $i$-form $\omega_{k}$ and a
$(p-i)$-form $\theta_{l}$ given by:%
\begin{equation}
\Omega_{k,l}^{(i),(p-i)}=\left\langle \omega_{k},\theta_{l}\right\rangle
=\underset{M_{p}}{\int}\omega_{k}\wedge\theta_{l},
\end{equation}
which defines a matrix of integers. The compactification of our 11D theory
therefore reduces to a theory of abelian differential forms:%
\begin{equation}
S_{(11-p)D}=\underset{i=0}{\overset{p}{\sum}}\underset{k_{i}%
=1}{\overset{b_{\text{cpct}}^{i}(M_{p})}{\sum}}\underset{l_{i}%
=1}{\overset{b_{\text{cpct}}^{p-i}(M_{p})}{\sum}}\frac{\Omega_{k_{i},l_{i}%
}^{(i),(p-i)}}{4\pi i}\underset{M_{11-p}}{\int}C_{k_{i}}^{(5-i)}\wedge
dC_{l_{i}}^{(5-p+i)}.
\end{equation}
Observe that this action breaks up into different non-interacting sectors. We
always have a theory of $(5-i)$-forms coupled to $(5-p+i)$-forms, but these do
not interact with the other sectors.\footnote{In a theory with additional bulk matter fields, there can be one-loop induced mixing
terms upon reduction to lower dimensions.} For this reason, we can treat these
contributions independently.

Note that we do not assume $M_{p}$ is compact. This means that the intersection pairings we generate, and
thus the resulting matrix of couplings need not be square matrices, and when they are square, they need not
have determinant one. For square matrices with $\det \Omega \neq 1$, the boundary theory is most appropriately
viewed as a relative quantum field theory in the sense of reference
\cite{Freed:2012bs} (see also \cite{Monnier:2017klz}).

\subsection{Examples}

Let us give a few examples to show how we recover various topological field
theories from this point of view. \ Isolating the contributions from the
middle degree forms, we get the following bulk theories:%
\begin{align}
S_{7D}  &  =\frac{\Omega_{IJ}^{(7D)}}{4\pi i}\int C_{(3)}^{I}\wedge
dC_{(3)}^{J}\label{CS7d}\\
S_{6D}  &  =\frac{\Omega_{IJ}^{(6D)}}{2\pi i}\int C_{(3)}^{I}\wedge
dB_{(2)}^{J}\label{BF6d}\\
S_{5D}  &  =\frac{\Omega_{IJ}^{(5D)}}{4\pi i}\int B_{(2)}^{I}\wedge
dB_{(2)}^{J}\label{CS5d}\\
S_{4D}  &  =\frac{\Omega_{IJ}^{(4D)}}{2\pi i}\int B_{(2)}^{I}\wedge
dA_{(1)}^{J}\label{BF4d}\\
S_{3D}  &  =\frac{\Omega_{IJ}^{(3D)}}{4\pi i}\int A_{(1)}^{I}\wedge
dA_{(1)}^{J}\label{CS3d}\\
S_{2D}  &  =\frac{\Omega_{IJ}^{(2D)}}{2\pi i}\int A_{(1)}^{I}\wedge
d\phi_{(0)}^{J}\label{BF2d}\\
S_{1D}  &  =\frac{\Omega_{IJ}^{(1D)}}{4\pi i}\int\phi_{(0)}^{I}\wedge
d\phi_{(0)}^{J} , \label{CS1d}%
\end{align}
in the obvious notation. There is a vast literature on nearly all of these theories,
and so we shall limit our discussion to a few general comments.

These bulk topological field theories
fall into two general subclasses, namely Chern-Simons-like
theories with potentials of the same degree and either
symmetric or anti-symmetric pairings, and BF-like theories with forms of different
degree. In all of these cases, we expect to realize
interesting edge mode dynamics, which in many cases can be understood from the
compactification of a chiral four-form in ten dimensions to the
lower-dimensional setting. Supersymmetry provides an additional extension of
these results and leads to an even broader class of lower-dimensional
theories.

One of the other lessons from string theory is that additional light degrees of freedom are
expected to emerge in limits where the compactification manifold develops
singularities. From this perspective, we can see that in many cases, we should
expect to realize both additional non-abelian structure and higher spin
currents in the boundary theory.

\subsubsection{BF-like theories}

We begin our discussion on effective topological field theories focusing
first on dimensions $6$, $4$ and $2$. Here, in general, we expect a BF-like
theory as in equations (\ref{BF6d}), (\ref{BF4d}), (\ref{BF2d}).

Such BF theories feature prominently in long distance limits of various high energy physics
systems and also play an important role in the description of gapped phases of matter. For example,
a four-dimensional action similar to (\ref{BF4d}) was instrumental in the
description of novel bosonic symmetry protected topological phases
\cite{Vishwanath:2012tq}.

The 6D BF-like theory appears to have not received as much
attention. Some details about this theory (and about BF theories of various
dimensions) can be found in Appendix A of \cite{Maldacena:2001ss}. It would be
very interesting to determine the resulting theory of edge modes. It would also likely
shed further light on the compactification of 6D SCFTs to five dimensions
(see e.g. \cite{Ganor:1996pc, Intriligator:1997pq, DelZotto:2017pti}).

\subsubsection{CS-like theories}

Consider next the Chern-Simons-like theories in which the differential
forms all have the same degree. This occurs when the number of spacetime dimensions
is odd.

The most familiar example in this class is given by abelian 3D Chern-Simons
theory (\ref{CS3d}), which as we have already remarked is helpful in the study of the
fractional quantum Hall effect \cite{Blok:1990mc, Frohlich:1991wb, Wen:1992vi, Zee:1996fe}.
Our interest in this paper has of course been the 7D generalization of this to three-forms.

Note that the theories in dimensions $7$ and $3$
have a symmetric matrix of couplings, as dictated by the intersection pairing
on the internal space. By contrast, the theories in dimensions $5$ and $1$
have an anti-symmetric matrix of couplings, again in accord with the structure of the
internal intersection pairing.

Finally, the 5D abelian Chern-Simons theory of two-forms in line
(\ref{CS5d}) has appeared both in the condensed matter and high energy theory literature.
For example, it appears in the low energy effective action of type IIB string theory
on the background $AdS_5 \times S^5$ \cite{Witten:1998wy, Aharony:1998qu, Maldacena:2001ss}. Alternative
applications of line (\ref{CS5d}) concern the study of both gapped phases of matter \cite{Kravec:2013pua}
and discrete symmetries of gauge \cite{Gaiotto:2014kfa}.

\section{Embedding in M-theory \label{sec:EMBED}}

In our discussion up to this point, we have deliberately phrased our entire
discussion in terms of a 7D topological field theory which is well-defined in
its own right. It is nevertheless of interest to see how the 7D
Chern-Simons-like theory we have been studying arises in compactifications of
M-theory. An added benefit of this approach is that we will automatically show
that there is a supersymmetric extension of our 7D theory.

Along these lines, we now take the 6D boundary theory to
be a Lorentzian signature manfold, so that the 7D bulk coordinate is an
additional spatial direction. Recall that we are interested in physical
systems with interacting degrees of freedom localized along the 6D boundary
which realize a chiral conformal field theory. At present, the only way to
construct examples of such theories are supersymmetric and involve embedding in string / M-theory / F-theory, the
first examples of this type being found in references \cite{Witten:1995zh,
Strominger:1995ac, Seiberg:1996qx}.

A helpful example to keep in mind is the
special case of M5-branes filling the first factor of
$\mathbb{R}^{5,1}\times\mathbb{R}_{\bot
}\times\mathbb{C}^{2}/\Gamma$ for $\Gamma$ a discrete subgroup of $SU(2)$.
We realize a conformal fixed point when the M5-branes all
sit at the orbifold singularity of $\mathbb{C}^{2}/\Gamma$ and a common point
of the $\mathbb{R}_{\bot}$ factor. This realizes the so-called class $\mathcal{S}_{\Gamma}$ conformal field
theories studied in references \cite{DelZotto:2014hpa, Gaiotto:2015usa,
Heckman:2016xdl}. We pass to the partial tensor branch of the
theory where effective strings have a tension by moving the M5-branes apart
from one another in the $\mathbb{R}_{\bot}$ direction. Observe that the
geometry $\mathbb{R}^{5,1}\times\mathbb{R}_{\bot}$ is seven-dimensional, so it
is natural to expect a bulk 7D theory to reside here which couples to the 6D
boundary defined by the M5-branes.

Let us now turn to the construction and study of this putative
7D massive supermultiplet which contains a three-form potential
and is decoupled from gravity. To see how this comes about,
it is actually helpful to proceed up to eight
dimensions, where supersymmetry has a chiral structure. Here,
we have the following 8D $\mathcal{N}=1$ massless supermultiplets
(see e.g. \cite{Strathdee:1986jr, Blumenhagen:2016rof}):%
\begin{align}
&  \text{8D Massless Supermultiplets:}\\
G_{\mathcal{N}=1}^{8D}  &  =1\cdot\lbrack2]+1\cdot\left[  \frac{3}{2}\right]
+2\cdot\lbrack1]+1\cdot\left[  \frac{1}{2}\right]  +1\cdot\left[  0\right]
+1\cdot\lbrack t_{2}]\\
S_{(3/2)}^{8D}  &  =1\cdot\left[  \frac{3}{2}\right]  +2\cdot\lbrack
1]+3\cdot\left[  \frac{1}{2}\right]  +2\cdot\lbrack0]+2\cdot\lbrack
t_{2}]+1\cdot\lbrack t_{3}]\\
V_{(1)}^{8D}  &  =1\cdot\left[  1\right]  +1\cdot\left[  \frac{1}{2}\right]
+2\cdot\lbrack0]
\end{align}
where the notation $[0]$, $[\frac{1}{2}]$, $[1]$, $[\frac{3}{2}]$, $[2]$
respectively refers to an 8D\ scalar, Weyl fermion, vector boson, gravitino
and graviton. The notation $[t_{p}]$ refers to a $p$-form potential. We also
have massive supermultiplets, where we denote massive fields by an
overline:%
\begin{align}
&  \text{8D\ Massive Supermultiplets:}\\
\overline{S^{8D}}_{(3/2)}  &  =1\cdot\overline{\left[  \frac{3}{2}\right]
}+ 1 \cdot \overline{[1]}+2\cdot\overline{\left[  \frac{1}{2}\right]  }+1\cdot\overline{
\lbrack0]}+1\cdot\overline{[t_{2}]}+1\cdot\overline{[t_{3}]}\\
\overline{V^{8D}}_{(1)}  &  =1\cdot\overline{\left[  1\right]  }%
+1\cdot\overline{\left[  \frac{1}{2}\right]  }+1\cdot\overline{[0]}.
\end{align}
In terms of the field content, we have the following relations:%
\begin{align}
G_{\mathcal{N}=2}^{8D}  &  =G_{\mathcal{N}=1}^{8D}+S_{(3/2)}^{8D}+2\cdot
V_{(1)}^{8D}\\
\overline{S^{8D}}_{(3/2)}  &  =S_{(3/2)}^{8D}\\
\overline{V^{8D}}_{(1)}  &  =V_{(1)}^{8D}.
\end{align}
The 8D $\mathcal{N}=2$ multiplet $G_{\mathcal{N}=2}^{8D}$
arises from M-theory compactified on a $T^{3}$.

Proceeding now to seven dimensions, we obtain the following irreducible 7D
supermultiplets:\footnote{We thank D.S. Park for helpful discussions.}
\begin{align}
G_{\mathcal{N}=1}^{7D}  &  =1\cdot\lbrack2]+2\cdot\left[  \frac{3}{2}\right]
+4\cdot\lbrack1]+4\cdot\left[  \frac{1}{2}\right]  +4\cdot\left[  0\right]
+1\cdot\lbrack t_{2}]\\
S_{(3/2)}^{7D}  &  =2\cdot\left[  \frac{3}{2}\right]  +4\cdot\lbrack
1]+8\cdot\left[  \frac{1}{2}\right]  +4\cdot\lbrack0]+3\cdot\lbrack
t_{2}]+1\cdot\lbrack t_{3}]\\
V_{(1)}^{7D}  &  =1\cdot\left[  1\right]  +2\cdot\left[  \frac{1}{2}\right]
+3\cdot\lbrack0].
\end{align}
The presence of the additional factor of two in the gravitino and Weyl spinors
has to do with the way we count degrees of freedom for our spinors;
In 8D we have a Weyl spinor, whereas in 7D we cannot have a Weyl spinor, and
instead impose an appropriate reality condition.

Note that we have not ``removed a vector multiplet'' from the gravitino multiplet
$S_{(3/2)}^{7D}$. The reason this is not correct to do can be seen either by directly
constructing the appropriate supermultiplet, or indirectly, by considering a
further reduction to 6D, where removing such a multiplet would make it
impossible to construct appropriate 6D supermultiplets. To see this
explicitly, we decompose $S_{(3/2)}^{7D}$ further into $6D$ fields:
\begin{align}
8D  &  \rightarrow7D\rightarrow6D\\
S_{(3/2)}^{8D}  &  \rightarrow S_{(3/2)}^{7D}\rightarrow2\cdot\left[  \frac
{3}{2}\right]  ^{+}+2\cdot\left[  \frac{3}{2}\right]  ^{-}+10\cdot\left[
\frac{1}{2}\right]  ^{+}+10 \cdot \left[  \frac{1}{2}\right]  ^{-}+8\cdot
\lbrack1]+8\cdot\lbrack0]+4\cdot\lbrack t_{2}].
\end{align}
And in six dimensions, we observe that we have the following $\mathcal{N} = (1,1)$ gravitino multiplets:
\begin{align}
&  6D\text{ Gravitino Supermultiplets:}\\
S_{(3/2)}^{+}  &  =\left[  \frac{3}{2}\right]  ^{+}+2\cdot\left[  1\right]
+4\cdot\left[  \frac{1}{2}\right]  ^{+}+1\cdot\left[  \frac{1}{2}\right]
^{-}+2\cdot\lbrack0]+2\cdot\lbrack t_{2}]^{+}\\
S_{(3/2)}^{-}  &  =\left[  \frac{3}{2}\right]  ^{-}+2\cdot\left[  1\right]
+4\cdot\left[  \frac{1}{2}\right]  ^{-}+1\cdot\left[  \frac{1}{2}\right]
^{+}+2\cdot\lbrack0]+2\cdot\lbrack t_{2}]^{-}.
\end{align}
So in other words, everything properly assembles into the following 6D
multiplets:%
\begin{align}
8D  &  \rightarrow7D\rightarrow6D\\
S_{(3/2)}^{8D}  &  \rightarrow S_{(3/2)}^{7D}\rightarrow2\cdot S_{(3/2)}%
^{+}+2\cdot S_{(3/2)}^{-}.
\end{align}
With this caveat dealt with, we now see that we have correctly identified a 7D
supermultiplet decoupled from the graviton:%
\begin{equation}
S_{(3/2)}^{7D}=2\cdot\left[  \frac{3}{2}\right]  +4\cdot\lbrack1]+8\cdot
\left[  \frac{1}{2}\right]  +4\cdot\lbrack0]+3\cdot\lbrack t_{2}]+[t_{3}].
\end{equation}

One might now ask whether it is actually consistent to consider such a
multiplet decoupled from gravity. Indeed, the presence of the Rarita-Schwinger
fields would seem to suggest that in any theory with a non-trivial S-matrix,
gravity must indeed be included.

But if we have a trivial S-matrix (though a non-trivial theory!),
we can consistently decouple gravity and simultaneously retain
our gravitino multiplet.\footnote{We thank M. Del Zotto for
discussions on this point.} To illustrate, observe that when we compactify
M-theory on a four-manifold, the action contains the following terms with the
three-form potential:%
\begin{equation}
\text{Vol}\left(  M_{4}\right)  \underset{7D}{\int}dC_{(3)}\wedge\ast
dC_{(3)}+ \frac{1}{4 \pi i} \underset{M_{4}}{\int} \frac{G_{(4)}}{2 \pi}\underset{7D}{\int}C_{(3)}\wedge
dC_{(3)}.
\end{equation}
In the case where we decouple gravity, Vol$\left(  M_{4}\right)  $ is quite
large, so in this sense the three-form is also ``decoupled''.
Additionally, however, we see that if there is a non-zero
four-form flux, as can happen if we have M5-branes present, then there is a
Chern-Simons-like coupling in seven dimensions. This places the
three-form potential in a gapped phase and in particular means that the
resulting supermultiplet is massive.

The fermions of the three-form multiplet decouple in a similar fashion. We
note that in the 11D supergravity action, there is a coupling of the schematic
form $\overline{\Psi}_{(3/2)}\cdot G_{(4)}\cdot\Psi_{(3/2)}$, so this flux
also generates a mass for the gravitinos and Weyl fermions of $S^{7D}_{(3/2)}$.
The four zero-form and four one-form potentials all develop masses through the
standard St\"uckelberg mechanism, and the kinetic terms:%
\begin{equation}
\left\vert d\phi_{(0)}+A_{(1)}\right\vert ^{2}+\left\vert dA_{(1)}\right\vert
^{2},
\end{equation}
with mass set by the compactification scale, just as for the three-form
potential. The final set of terms involving the two-form potentials proceed in
analogous fashion. Observe that the magnetic dual of the two-form is a
three-form, so we can alternatively just write additional Chern-Simons like
couplings to decouple these modes as well. All told, then, we see that the
local excitations decouple, the only remnant in the 7D theory being a
topological sector.

One can also consider coupling the 7D gravitino multiplet to the 7D vector
multiplet.\ This proceeds, for example, through the topological term:
\begin{equation}
\mu_{7D}\underset{7D}{\int}C_{(3)}\wedge\text{Tr}(F\wedge F),
\end{equation}
where $F$ is the two-form field strength and $\mu_{7D}$ is the
\textquotedblleft lift\textquotedblright\ of the 6D Green-Schwarz
coupling \cite{Green:1984bx, Sagnotti:1992qw, Sadov:1996zm, Morrison:2012np}
to seven dimensions.

It is also possible to couple our supersymmetric theory to defects carrying
propagating massless degrees of freedom. Observe that in this case, we need to
impose suitable boundary conditions which will break half of the
supersymmetry, but allow us to couple to appropriate supersymmetric edge modes.

In the above we have only sketched the main elements of the 7D supersymmetric
theory. It would of course be most instructive to fill in these details in
future work.

\section{IIB as an Edge Mode \label{sec:Ftheory}}

Motivated by the success of embedding our 7D topological theory in a
supersymmetric compactification of M-theory, it is natural to ask whether a
similar structure also holds for IIB string theory. Indeed, IIB\ supergravity
contains a chiral four-form potential with self-dual field strength and there are
notorious subtleties in writing an off-shell 10D action.

From our present perspective, the issue is in some sense forced once we demand
proper quantization of fluxes in the quantum theory. In the special case
where we restrict to level one for the chiral four-form (as happens for the
standard IIB supergravity action) then we have an invertible theory in the sense of \cite{Freed:2012bs} so we expect to be able to
decouple the bulk and boundary. Nevertheless, such an answer is rather unsatisfying because one could ask what happens at other
``levels'' of the chiral four-form, namely if we choose a different normalization for the two-point function. It is also
disturbing that certain quantum questions cannot be easily accessed due to the absence of an off-shell action.

With this in mind, it is tempting to extend the standard 10D spacetime by an additional spatial
direction to write an 11D topological theory of five-forms \cite{Belov:2006jd, Belov:2006xj},
much as we already did in section \ref{sec:CPCT}. We have already seen the utility of this
perspective in our geometric unification of lower-dimensional topological theories.

Having gone to eleven dimensions, it is now irresistible to try and connect
this with the 12D geometric formulation of F-theory \cite{Vafa:1996xn}. Our
aim in this section will be to present some suggestive --though conjectural--
aspects of how to view all of IIB\ superstring theory as an edge
mode of a bulk $10+2$-dimensional topological theory.\footnote{We thank C. Vafa for
an inspirational discussion.} In some sense this
resurrects the original formulation of F-theory as genuinely living in
$10+2$ dimensions rather than \textquotedblleft merely\textquotedblright\ as a
formal device for constructing non-perturbative 10D vacua. It also points
the way to the construction of new self-consistent string vacua.

To accomplish this, we shall abandon from the start the notion of having
propagating degrees of freedom in all twelve dimensions. Indeed, propagating
degrees of freedom with two time directions leads to severe pathologies with
causality. Rather, we shall immediately restrict attention to edge modes
localized on a $9+1$-dimensional subspace. Our goal will be to see whether
this can be fit together self-consistently.

To motivate our conjectural formulation of IIB\ as an edge mode, we shall
proceed in a \textquotedblleft bottom up\textquotedblright\ fashion, piecing
together various consistency conditions. These will include:

1)\ We attempt to quantize the IIB\ self-dual five-form flux by viewing the
chiral four-form potential as an edge mode coupled to a bulk topological theory.

2) We require the bulk theory to preserve supersymmetry, namely we must be
able to assemble bulk supermultiplets.

3)\ We must maintain the known $SL(2,\mathbb{Z})$ covariance central to the
formulation of IIB\ and its geometrization in F-theory.

Let us now proceed to enforce conditions 1) - 3). Along these lines, we briefly
recall that the Ramond-Ramond sector of IIB\ string theory consists of a
chiral four-form $c_{(4)}$, a two-form $c_{(2)}$, and a zero-form $c_{(0)}$.
Continuing with our general philosophy, we shall at first attempt to lift this
to an 11D space with 10D boundary. In our 11D space we introduce bulk
higher-form potentials $C_{(5)}$, $C_{(3)}$ and $C_{(1)}$ so that the edge
modes amount to degrees of freedom which cannot be gauged away, namely
$C_{(p)}=dc_{(p-1)}$ on the boundary.

We immediately face an obstacle because the 11D supergravity spectrum in
$10+1$ dimensions does not admit such degrees of freedom!\ There is, however,
a loophole if we proceed to $10+2$ dimensions, namely
ten spatial and two temporal dimensions. Supergravity multiplets in a 12D
spacetime with this exotic signature have been considered previously, for
example in references \cite{Castellani:1982ke, Bergshoeff:1982az,
Blencowe:1988sk, Bars:1996dz, Hewson:1996yh, Nishino:1997gq, Nishino:1997sw,
Hewson:1997wv}. The key point is that in $10+2$ dimensions, one can impose a
Majorana-Weyl condition on spinors which in turn allows for a match between
bosonic and fermionic degrees of freedom. The supersymmetry algebra in $10+2$
dimensions has been studied in references \cite{Bars:1996dz, Nishino:1997sw},
and a proposed chiral supergravity theory has been constructed in
\cite{Blencowe:1988sk, Nishino:1997gq, Nishino:1997sw} (for a different
perspective, see e.g. \cite{Choi:2014vya, Choi:2015gia}). The general
conclusion is that one ought to consider a theory with a four-form and
two-forms, as well as a pair of null one-forms, $\mu_{+}$ and $\mu_{-}$.

Before proceeding to how this matches with the form content we have, let us
briefly explain how this earlier work makes contact with the more geometric
formulation of F-theory on an elliptically fibered Calabi-Yau with section which
has become the de facto standard in modern practice.
Along these lines, let us suppose that we have a standard F-theory background
of the form $\mathbb{R}^{d-1,1}\times CY_{n}$, where $CY_{n}$ is an
elliptically fibered Calabi-Yau $n$-fold with base $\mathcal{B}_{n-1}$ a K\"{a}hler manifold
of complex dimension $(n-1)$ so that the total spacetime dimension is
$10=d+2(n-1)$. The existence of a section means that there is an embedding of
the 10D spacetime $\mathbb{R}^{d-1,1}\times \mathcal{B}_{n-1}$ in the 12D\ geometry. As
such, we also have a local normal coordinate $z \in \mathcal{O}(-K_{\mathcal{B}_{n-1}})$,
and corresponding infinitessimal one-form $dz$. This $dz$ can in turn
be decomposed into real and imaginary parts, and thus we get a pair of real
one-forms, namely the Euclidean analogs of $\mu_{+}$ and $\mu_{-}$ presented
above. Extending to the 12D total space, we can also construct a connection
which parameterizes the breaking of $SO(10,2)$ to $SO(9,1)$:%
\begin{equation}
\mu_{(1)}=\mu_{+} \cdot d\mu_{-}. \label{mumumu}%
\end{equation}
Here, the ``$\cdot$'' refers to the contraction of one-form indices.\footnote{For
the perhaps more familiar case of breaking patterns of a compact Lie group, we would write
$\mu_{(1)} = -i g^{\dag} d g$, where $g$ is a generator of the group.
Note that for a $U(1)$ gauge theory with $g = \exp(i \theta)$, the
formal one-form reduces to $d \theta$.} This relation appears (in a slightly different form)
in reference \cite{Schwarz:1983qr} (see also \cite{Nishino:1997sw}),
for example.\footnote{In reference \cite{Nishino:1997sw}, the $SL(2,\mathbb{Z})$ duality relation of
IIB\ supergravity is imposed \textquotedblleft by hand\textquotedblright\ at
the classical 12D level, rather than as an emergent property from
compactification on a $T^{2}$.}

Continuing in Euclidean signature for our two extra dimensions,
the local geometry near the 10D spacetime is the Cartesian product of a cylinder $\mathcal{C}$
with the 10D spacetime. Topologically, $\mathcal{C} = S^1 \times \mathbb{R}$. The local profile of the metric
is in the conformal class:
\begin{equation} \label{warped}
ds_{Euc}^{2} = d \theta^{2} + d r^2,
\end{equation}
so that $\theta$ is a periodic local coordinate,
and $r$ parameterizes the radial profile of the cylinder.
Globally, we of course complete this local presentation of the metric
to construct an elliptic fibration over the 10D spacetime,
with metric in the conformal class:
\begin{equation}\label{EUCLID}
ds_{E}^{2}=\left(  dx+\tau(x_{10D}) dy\right)  \left(  dx+\overline{\tau}(x_{10D}) dy\right)  ,
\end{equation}
where the complex structure of the elliptic curve is identified with the
combination of IIB\ supergravity modes:%
\begin{equation}
\tau=c_{(0)}+ie^{-\phi}.
\end{equation}
This elliptic fibration is central to the geometrization
of $SL(2,\mathbb{Z})$ duality in IIB string theory.

Let us now turn to the Lorentzian signature version, as dictated by supersymmetry in $10 + 2$ dimensions.
We have already presented evidence that to properly quantize the 10D chiral four-form, we ought to extend our spacetime by one additional
spatial direction, i.e., the $r$ coordinate of line (\ref{warped}).
This means that if we insist on a supersymmetric theory in $(10,2)$ signature,
we need to analytically continue $\theta \mapsto - i \theta$.
So for us, the local profile of the metric in line (\ref{warped})
is instead:
\begin{equation}\label{LorWarped}
ds_{Lor}^{2} = - d \theta^{2} + dr^2.
\end{equation}
The presence of an additional timelike direction means that a reduction
along a null direction takes us from a $10+2$-dimensional
spacetime to a $9+1$-dimensional spacetime.

At this point, something awkward appears to have happened.
On the one hand, we have argued based on flux quantization considerations
and supersymmetry that we need to add Lorentzian directions of signature $(1,1)$.
On the other hand, the $SL(2,\mathbb{Z})$ duality group of IIB is most
naturally formulated in terms of a Euclidean $T^2$, and simply analytically continuing
to Lorentzian signature fails to retain this structure.\footnote{The main point is that
it is not possible to define a family of Lorentzian signature torii in which the analog of $\tau$
smoothly varies. We thank D.R. Morrison for helpful comments on this point.}
Again taking our cue from the original paper \cite{Vafa:1996xn},
we note that the Narain lattice for strings on a Euclidean $T^2$ has signature $(2,2)$ so it is not
inconceivable that a $(1,1)$ signature spacetime could still replicate these features.

\begin{figure}[t!]
\centering
\includegraphics[trim = 0cm 4.75cm 0cm 4.75cm, clip=true, scale=0.5]{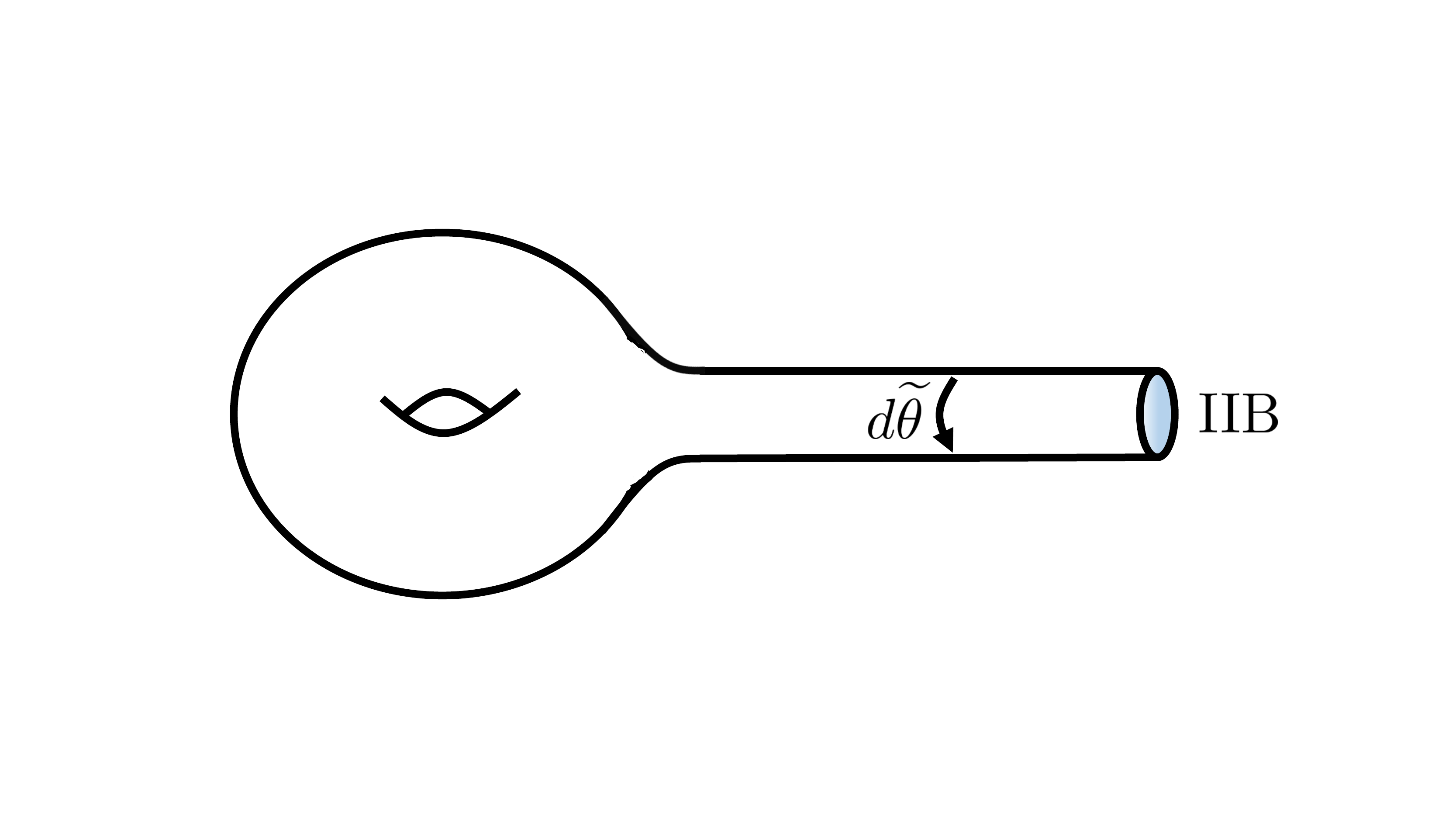}
\caption{Depiction of the 12D geometry associated with F-theory and IIB as an edge mode
in the special case where the axio-dilaton is constant.
Here, we have passed from the $(10,2)$ signature spacetime to one in which a dual
winding coordinate $\widetilde{\theta}$ appears. This takes us to a dualized spacetime
of signature $(11,1)$ and in which the standard elliptic fiber of F-theory appears geometrically.}%
\label{IIBedge}%
\end{figure}

To put this on a firmer footing, we now observe that by construction, we do not allow any localized excitations in the two extra directions, this being the point of dealing with a bulk 12D topological theory in the first place. Nothing,
however, disallows extended objects to wrap around compact cycles of the geometry.
In particular, returning to our discussion of the Lorentzian cylinder geometry of line (\ref{LorWarped}), we see
that we can wrap an extended object along the timelike circle parameterized by the $\theta$ coordinate. The winding number along
this circle has a Fourier transform which yields a dual periodic coordinate $\widetilde{\theta}$. From this perspective, there is again a
Euclidean signature torus lurking, but in comparison with line (\ref{EUCLID}), it will involve
a dual coordinate $\widetilde{x}$:
\begin{equation}
ds_{\widetilde{E}}^{2}=\left(  d \widetilde{x} + \tau(x_{10D}) dy\right)  \left(  d \widetilde{x} + \overline{\tau}(x_{10D}) dy\right).
\end{equation}
This sort of mixing of momentum and winding is a
feature of geometric approaches to T-duality such as double field theory
\cite{Siegel:1993th, Siegel:1993xq, Hull:2009mi}.
In some sense our discussion is less ambitious since we
restrict to a topological sector from the start. We thus have in our extra $1+1$-dimensions
a pair of null directions along which we can compactify on lightlike circles. Even though nothing propagates along them,
extended objects can wrap the circles, and the consistent description of such degrees of freedom leads to a Euclidean signature torus,
with complex structure the axio-dilaton of IIB strings. See figure \ref{IIBedge} for a depiction of the geometry in the
special case in which we have a constant elliptic fibration.

A satisfying byproduct of this proposal is that it helps to clarify the role of the elliptic fiber in M-theory / F-theory duality. Recall that in
the standard picture, M-theory on the background $CY_n$ is also described by F-theory on $S^1 \times CY_n$, where $CY_n$ denotes the ``same''
elliptically fibered Calabi-Yau $n$-fold. In this process, one is supposed to collapse the M-theory elliptic fiber to zero size to reach the F-theory limit. In the lower-dimensional context of IIA / IIB T-duality, one can consider similar limits, modulo the caveat that we interchange momentum and winding degrees of freedom. Here, we see that a similar consideration seems to hold, and in some sense is required to simultaneously satisfy our
earlier conditions 1) - 3) mentioned at the beginning of this section. Precisely because we can interchangeably work in terms of either a
$10+2$-dimensional spacetime, or one in which we trade a position coordinate for a winding coordinate to reach a dualized
$11+1$-dimensional spacetime, we see that the procedure of dimensional reduction
also carries over. In particular, a reduction of differential forms
along the circle with coordinate $\widetilde{\theta}$
proceeds as in a standard Kaluza-Klein compactification.

With the geometry of the spacetime dealt with, let us now return to our proposed mode content,
now lifted to $10+2$ dimensions. We have so far identified a five-form, three-form and one-form,
namely $C_{(5)}$, $C_{(3)}$, and $C_{(1)}$. Additionally, taking
our cue from earlier work on supersymmetry in $10+2$ dimensions, we impose an
anti-self-duality condition on the six-form field strength
$F_{(6)} = -\ast_{12D} F_{(6)}$. Given that our motivation was to properly quantize the
fluxes of the IIB theory, and we have now introduced a self-duality constraint
in 12D, one might ask whether it is necessary to proceed up to 13 dimensions, perhaps
along the lines of S-theory \cite{Bars:1996ab}. There is no reason to do so, at least from quantization
considerations. The reason is that because of the presence of a distinguished null one-form, the way we read off
the physical brane spectrum is different. Indeed, as we shortly explain, the
physical degrees of freedom associated with our five-form potential require either contraction or wedging our five-form
to a four-form or six-form, respectively. As such, the physical degrees of freedom are
associated with a $2+2$-brane and $4+2$-brane, so we need not directly quantize the
corresponding six-form flux units (since we have no $3+2$-branes to speak of). Indeed,
turning the discussion around, we see that the absence of integral six-form fluxes
actually \textit{eliminates} the possibility of a $3+2$-brane,
leaving us only with objects such as a $2+2$-brane or $4+2$-brane!

Comparing with the earlier supersymmetry literature in $10+2$ dimensions
\cite{Castellani:1982ke, Bergshoeff:1982az, Blencowe:1988sk, Bars:1996dz, Hewson:1996yh,
Nishino:1997gq, Nishino:1997sw, Hewson:1997wv},  we now ask how we can recover a four-form potential
and a pair of two-form potentials, as well as the axio-dilaton. This all falls into place
upon contracting our $p$-forms with the one-forms $\mu_+$ and $\mu_-$, namely we write:\footnote{Here,
we are working in the $(11,1)$ signature geometry with a winding coordinate
so as to make $SL(2,\mathbb{Z})$ duality manifest.}
\begin{equation}
\mu_{\pm} \cdot C_{(p)} \equiv C^{(\pm)}_{(p - 1)}.
\end{equation}
Note that this automatically generates covariant transformation rules for the
two-form potentials and the scalars in the 10D IIB theory, and the
anti-self-duality relation projects us onto an $SL(2,\mathbb{Z})$ singlet for the chiral four-form potential.

A closely related point is that (as is well-known from M- / F-theory duality)
the periods of our potentials along circles
produces the expected IIB\ mode content. For example, the three-form reduces
to the NS\ and RR\ two-forms, and transform as a doublet of $SL(2,\mathbb{Z})$. The periods of the one-form
potential $C_{(1)}$ along the two one-cycles produce a pair of scalars, and ratios of
appropriate linear combinations reduces to the axio-dilaton $\tau$.
Finally, the periods of the five-form are related
by duality of its field strength, namely $F_{(6)}=-\ast_{12D}F_{(6)}$, with
$F_{(6)}=dC_{(5)}$, so the boundary four-form transforms as an $SL(2,\mathbb{Z})$
singlet.

So far, we have focused on the RR sector, and have already seen a partial
unification with some parts of the NS\ sector of the 10D\ boundary. To round
out our discussion, we now turn to the 12D origin for the remaining
NS\ degrees of freedom, namely the 10D\ metric. Since we have
introduced explicit one-forms $\mu_{+}$ and $\mu_{-}$, we expect the graviton
to always have a formally infinite mass in two of the twelve directions.
The gauge redundancy for a massless graviton:%
\begin{equation}
g_{AB}\rightarrow g_{AB} + \nabla_{A}C_{B}+\nabla_{B}C_{A},
\end{equation}
also means that we can alternatively package our mode content in terms of a
10D graviton, and a one-form $C_{(1)}$, the one introduced earlier from a
bottom up perspective. By the same token, we expect a partially massive
gravitino in $10+2$ dimensions to fill out the necessary fermionic degrees of freedom.

Let us now turn to the sense in which our system is actually in a gapped
phase. Starting from the 11D action:%
\begin{equation}
\frac{1}{4\pi i}\underset{11D}{\int}C_{(5)}\wedge dC_{(5)}\text{,}
\label{11Dagain}%
\end{equation}
we can extend to 12D using our privileged one-form $\mu_{(1)}$ of line
(\ref{mumumu}). By the same token, we can also introduce a related BF-like
term to produce a mass gap for the one-form $C_{(1)}$ and its pairing to the
three-form $C_{(3)}$. Writing out all proposed couplings to the higher-form
potentials, we have:%
\begin{align}
S_{12D,\text{top}}  &  =\frac{1}{4\pi }\underset{12D}{\int}\mu_{(1)}\wedge
C_{(5)}\wedge dC_{(5)}+\frac{1}{4\pi }\underset{12D}{\int}\left(  \mu
_{(1)}\cdot C_{(3)}\right)  \wedge dC_{(9)} \label{topo}\\
&  +\frac{1}{4\pi }\underset{12D}{\int}\left(  \mu_{(1)}\cdot C_{(5)}\right)
\wedge G_{(4)}\wedge G_{(4)}, \label{boto}
\end{align}
where $G_{(4)}$ is the four-form field strength for the three-form potential,
and we have used the magnetic dual of $C_{(1)}$, namely the nine-form $C_{(9)}$.
The coupling of the second line has appeared in the F-theory literature before
\cite{Choi:2014vya, Choi:2015gia, Ferrara:1996wv, Donagi:2008kj}. Observe that
we have dropped pre-factors of ``$i$'' because we have two Lorentzian directions. Finally,
note that the only appearance of the metric occurs in the privileged null direction via contraction
with $\mu_{(1)}$. This dependence is sufficiently mild that we again obtain a trivial stress energy tensor,
as required for a topological theory of this type. As far as we are aware,
the couplings of line (\ref{topo}) have not been
considered before.

We can also include a partially massive graviton, along the lines
of reference \cite{deRham:2010kj}. Such theories have appeared in string
theory and related holographic setups, see e.g. \cite{Vegh:2013sk,
Blake:2013bqa, Blake:2013owa}. Let us note that in
massive gravity, there are various concerns about superluminal
propagation. Here, however, we are taking a rather different limit from what
is typically discussed in phenomenological applications: For us, the mass in the
decoupled directions is formally infinite, rather than small.
An additional comment is that in the related context of warped
compactification, a graviton zero mode can be localized
on a lower-dimensional spacetime \cite{Karch:2000ct}. This is rather suggestive considering that
in F-theory models, the base of the model has positive curvature, and thus
the curvature is negative in the directions normal to the base inside the Calabi-Yau.

Returning to the topological couplings of our 12D action, note that in the 10D
boundary, the first term reduces to a kinetic term for a chiral four-form, and
the second term provides a St\"{u}ckelberg mass for the RR\ and
NS\ two-forms. Indeed, in F-theory backgrounds with non-trivial
$SL(2,\mathbb{Z})$ monodromy, these two-forms are generically projected out,
receiving a mass due to such interactions. Line (\ref{boto}) reduces to the
well-known IIB\ coupling $c_{(4)}\wedge h_{(3)}^{RR}\wedge h_{(3)}^{NS}$,
which could in principle receive additional one-loop
contributions from integrating out components of the massive gravitino.

It is also tempting to extend each
stringy excitation of the 10D boundary by a $1+1$-dimensional direction in 12D, as for example in
\cite{Bars:1996dz, Hewson:1996yh, Nishino:1997gq, Nishino:1997sw, Hewson:1997wv}.
Following our topological route, we extend the F1- and D1-branes to a topological $2+2$-brane which
couples to the four-form obtained through the contraction $\mu
_{(1)}\cdot C_{(5)}$. Some aspects of $2+2$-dimensional branes have recently
been studied for example in \cite{Linch:2015fca}, and for earlier work, see
e.g. \cite{Ooguri:1990ww, Ooguri:1991fp, Ooguri:1991ie, Kutasov:1996zm,
Kutasov:1996vh}. We can also extend a $3+1$-dimensional D3-brane to a
$4+2$-dimensional topological brane which couples to the
six-form $\mu_{(1)}\wedge C_{(5)}$.

This also points the way to constructing new string vacua. For
example, though we have for the most part concentrated on the case of unit
coefficient in line (\ref{11Dagain}), it is quite tempting to broaden our
horizons to more general couplings of the form:%
\begin{equation}
\frac{N}{4\pi i}\underset{11D}{\int}C_{(5)}\wedge dC_{(5)}\text{.}%
\end{equation}
There is a simple interpretation of this as winding modes along a
timelike circle of the 12D geometry. Indeed, in an early speculative attempt
to mimic the M-theory interpretation of D0-branes as momentum along a circle,
reference \cite{Tseytlin:1996ne} sought to interpret IIB\ D-instantons (namely
$D_{-1}$-branes) in terms of ``momentum'' along a circle of a 12D
spacetime. Here, we see that a background condensate of D-instantons produces
a corresponding shift in the level of the five-form theory. In this
formulation, we identify $N-1$ with a winding number for the metric, since the
case of no D-instantons ought to correspond to the case of the ``usual''
IIB\ background with canonically normalized two-point function for the chiral four-form in ten dimensions. This provides a
somewhat complementary motivation for IIB matrix models, and it would be interesting
to revisit earlier proposals such as \cite{Ishibashi:1996xs, Aoki:1998vn, Aoki:1998bq}.
Departing one step further, it is tempting to consider 12D geometries
without an elliptic fibration.

Apparently, then, the natural setting for many of our considerations resides
in a $10+2$-dimensional bulk topological theory. This appears to be compatible
with the conditions of supersymmetry, $SL(2,\mathbb{Z})$ duality, and also evades
the pathologies of two-time physics and potential issues with massive gravity.
We leave more detailed checks of this proposal for future work.

Summarizing, the 12D topological interpretation of superstrings is intriguing.

\section{Conclusions \label{sec:CONC}}

In this paper we have proposed a six-dimensional generalization of the
fractional quantum Hall effect which makes use of the correspondence
between a bulk 7D topological field theory and a 6D theory of edge modes. The
bulk is given by a Chern-Simons like 7D theory of three-forms, and the edge
mode theory is that of free anti-chiral two-forms. In the presence of a large
background magnetic four-form flux, there is a direct analog
with the usual fractional quantum Hall effect. We determined the leading
order behavior of the analogous Laughlin wavefunction in various limits, and
have also explained how this higher-dimensional starting point provides a
unifying perspective on several lower-dimensional systems involving a bulk
topological field theory coupled to edge modes. This
7D theory embeds in a limit of M-theory decoupled from gravity, and this in turn
motivates a speculative conjecture on the interpretation of F-theory as a
topological theory in $10+2$ dimensions coupled to $9+1$-dimensional edge
modes associated with IIB\ strings. In the remainder of this section we
discuss some avenues for future investigation.

There is by now a nearly complete list of supersymmetric 6D CFTs. One of the
original motivations for this work was to better understand the topological structure
of these fascinating theories. It is therefore quite tempting to ask whether we
can develop a supersymmetric version of our analysis. For some discussion of a
supersymmetric version of the fractional quantum Hall effect in two
dimensions, see for example \cite{Tong:2015xaa, Vafa:2015euh}.

In this work focused on the simplest situation where there is a
single connected component to the boundary. In lower dimensions, it is often
quite fruitful to consider additional disconnected components to the boundary.
This in turn is intimately connected with the spectrum of non-local operators
in the theory.

Another thread of our analysis is the unifying framework it provides for
generating a rich class of lower-dimensional phenomena. We can already
anticipate that various integer couplings can in many cases be recast as
topological properties of a compactification manifold. It would be exciting to
use this perspective to develop a systematic analysis of these possibilities.

One of the original motivations for this work was to better understand the
topological sector of little string theories (for early constructions see e.g. \cite{Witten:1995zh,Aspinwall:1996vc,Aspinwall:1997ye,Seiberg:1997zk,Intriligator:1997dh, Hanany:1997gh, Brunner:1997gf}),
in which the pairing $\Omega_{IJ}$ is no longer invertible since it has a zero eigenvalue
\cite{Bhardwaj:2015xxa, Bhardwaj:2015oru}. Such little string
theories are non-local, and thus not controlled by the standard
axioms of local quantum field theory. It is natural to ask whether a
suitable 7D theory exists with edge modes given by a
little string theory.

We have also seen that our 7D topological theory admits a supersymmetric
extension which embeds as a decoupling limit of the physical M-theory.
Topological M-theory provides a unified framework for understanding
various aspects of topological strings \cite{Dijkgraaf:2004te} (see also
\cite{Gerasimov:2004yx, Nekrasov:2004vv}).
This theory is formulated in terms of an abelian three-form, but
instead involves compactifying on a manifold of
$G_{2}$ metric holonomy with action the integral of $C_{(3)}\wedge\ast
_{7D}C_{(3)}$, where the metric is itself constructed from the associative
three-form \cite{Hitchin:2000jd, Hitchin:2001rw}. Adding such a term to a
theory of a three-form gauge potential also leads to a mass gap. It would thus be
interesting to compare the IR behavior of these two 7D theories.

Proceeding up to $10+2$ dimensions, we presented some tantalizing hints of
an interpretation of F-theory as a bulk topological theory. Though quite
conjectural, it already provides a novel starting point for realizing new
classes of string vacua. It would clearly be very interesting to provide
further evidence for this proposal. A particularly fascinating
aspect of such a correspondence is the emergence of the IIB\ graviton as an
edge mode excitation.

\section*{Acknowledgements}

We thank F. Apruzzi, M. Del Zotto, T.T. Dumitrescu, O.J. Ganor, D.R. Morrison, D.S. Park and C. Vafa for
helpful discussion. We also thank O.J. Ganor and D.R. Morrison for comments on an earlier draft.
JJH thanks the 2016 and 2017 Summer workshops at the Simons Center for Geometry and Physics
as well as the Aspen Center for Physics Winter Conference in 2017
on Superconformal Field Theories in $d\geq4$, NSF grant PHY-1066293, for
hospitality during part of this work. The work of JJH is supported by NSF
CAREER grant PHY-1452037. LT thanks UNC\ Chapel Hill and the ITS\ at CUNY\ for
hospitality during this work. The work of LT is supported by VR grant
\#2014-5517 and by the \textquotedblleft Geometry and Physics" grant from Knut
and Alice Wallenberg Foundation.

\newpage

\bibliographystyle{utphys}

\bibliography{HighFQHE}

\end{document}